\documentclass[useAMS,usenatbib]{mn2e}

\title[Pilot Observations for MALT-45]
  {Pilot Observations for MALT-45: A Galactic Plane Survey at 7\,mm}
\author[C. H. Jordan et al.]
  {C.~H.~Jordan,$^{1,2}$\thanks{E-mail: christopher.jordan@jcu.edu.au}
  A.~J.~Walsh,$^1$
  V.~Lowe,$^{2,3}$
  N.~Lo,$^4$
  C.~R.~Purcell,$^5$\newauthor
  M.~A.~Voronkov$^2$
  and S.~Longmore$^6$ \\
  $^1$Centre for Astronomy, School of Engineering and Physical Sciences, James Cook University, Townsville, QLD 4814, Australia\\
  $^2$Australia Telescope National Facility, CSIRO Astronomy and Space Science, PO Box 76, Epping, NSW 1710, Australia\\
  $^3$School of Physics, University of New South Wales, Sydney, NSW 2052, Australia\\
  $^4$Departamento de Astronom\'ia, Universidad de Chile, Camino El Observatorio 1515, Las Condes, Santiago, Casilla 36-D, Chile\\
  $^5$School of Physics, University of Sydney, Sydney, NSW 2006, Australia\\
  $^6$ESO Headquarters, Karl-Schwarzschild-Str. 2, 85748 Garching bei M\"{u}nchen, Germany
}
\date{Accepted 2012 November 2. Received 2012 October 26; in original form 2012 August 28}

\pagerange{\pageref{firstpage}--\pageref{lastpage}} \pubyear{2012}

\def\LaTeX{L\kern-.36em\raise.3ex\hbox{a}\kern-.15em
    T\kern-.1667em\lower.7ex\hbox{E}\kern-.125emX}

\usepackage{subfigure}
\usepackage{graphicx}
\usepackage{xspace}

\usepackage{color}
\definecolor{highlight}{rgb}{1,0,0}

\newcommand{\kms}{\mbox{km\,s$^{-1}$}\xspace}
\newcommand{\ms}{\mbox{m\,s$^{-1}$}\xspace}

\newcommand{\cs}{CS\xspace}
\newcommand{\choh}{CH$_3$OH\xspace}
\newcommand{\sio}{SiO\xspace}
\newcommand{\nh}{NH$_3$\xspace}
\newcommand{\water}{H$_2$O\xspace}
\newcommand{\co}{CO\xspace}

\newcommand{\hi}{H\,{\sevensize I}\xspace}
\newcommand{\hii}{H\,{\sevensize II}\xspace}

\begin{document}

\label{firstpage}

\maketitle

\begin{abstract}
We introduce the MALT-45 (Millimetre Astronomer's Legacy Team - 45\,GHz) Galactic plane survey and describe pilot survey results with the Australia Telescope Compact Array (ATCA). The pilot survey was conducted to test the instrumentation and observational technique of MALT-45, before commencing the full survey. We mapped two half-square degree regions within the southern Galactic plane around the G333 giant molecular cloud, using fast mosaic mapping. Using the new Compact Array Broadband Backend (CABB) on the ATCA, we were able to observe two 2048\,MHz spectral windows, centred on frequencies 43.2 and 48.2\,GHz. Although only a coarse spectral resolution of around 7\,\kms was available to us, we detect widespread, extended emission in the \cs (1--0) ground state transition. We also detect eight Class I \choh masers at 44\,GHz and three \sio masers in vibrationally excited (1--0) transitions. We also detect the H53{\footnotesize{$\alpha$}} radio recombination line, non-vibrationally excited \sio (1--0) and emission in the \choh 1$_1$--0$_0$ A$^{+}$ line.
\end{abstract}


\section{Introduction}

High mass stars have a pivotal influence on the structure of our Galaxy, particularly at the start and end of their lives. Young high mass stars are responsible for the evolution of molecular clouds, for example, through their powerful, ionising winds and the copious amounts of energy they inject into the interstellar medium (ISM) \citep{zinnecker07}. The energy a high mass star exhibits over its lifetime can rival that of a supernova explosion. Finally, high mass stars end their lives as supernovae, again dramatically altering the structures of the Galaxy, transforming the ISM and dispersing heavy elements.

High mass star formation (HMSF) is a poorly understood process due to the rapid evolution of high mass stars, rarity, and high levels of dust extinction \citep{zinnecker07}. HMSF does not occur in isolation, thus making it difficult to disentangle mechanisms involved with HMSF from other sites of star formation in the vicinity. HMSF is known to occur within dense regions of giant molecular clouds (GMCs; eg. \citealt{lo11}). In order to comprehend the earliest stages of HMSF, a catalogue of their regions must be identified and mapped through high density tracers.

Previous surveys have been productive in identifying many sites of HMSF, but emphasise signposts that necessarily pick out only a subsection of the various stages. For example, methanol maser surveys (eg. \citealt{walsh98,green09}) pick out hot cores and young ultracompact \hii regions; infrared dark clouds (IRDCs; eg. \citealt*{rathborne06}) are only seen when they are close to us, and projected against a bright infrared background from the Galaxy and may not always identify sites of high mass star formation (\citealp{kauffmann10}); radio continuum sources are only found once a high mass star has had time to reach the main sequence and push its ionising wind through the surrounding material (eg. \citealt{purcell10}).

Untargeted Galactic plane surveys not only give us an opportunity to address these concerns, but they also allow us to study aspects of HMSF on a Galaxy-wide scale, with higher resolution and sensitivity than is achievable in external galaxies. For example, we can study the structure of the Galaxy through \co surveys (eg. \citealt{dame08}) and \hi surveys (eg. \citealt{mcclure-griffiths05}). The nature of \co and \hi mean that these tracers typically pick out low density gas ($10^{1-2}$\,cm$^{-3}$) that is not directly related to HMSF. These tracers are also sensitive to inter-spiral arm gas, making it difficult to identify spiral structure in the Galaxy. Higher density gas tracers, such as \nh and HC$_3$N, observed as part of the \water southern Galactic Plane Survey (HOPS; \citealt{walsh08,walsh11}), do trace gas directly associated with HMSF and may more clearly trace spiral structure. However, HOPS is a shallow survey, sensitive to typical HMSF regions within a few kpc. In order to study the spiral structure of the Galaxy, as well as the gas associated with HMSF, we have devised the MALT-45 survey. This paper introduces the MALT-45 survey design and outlines the autocorrelation results of pilot survey. MALT is the Millimetre Astronomer's Legacy Team - the result of a successful meeting of star formation astronomers to collaborate on major southern hemisphere millimetre surveys. MALT currently consists of the MALT-90 \citep{foster11} and MALT-45 surveys.

\section{MALT-45 Design}
MALT-45 aims to provide a comprehensive survey of star formation tracers at 7\,mm ($\sim$\,45\,GHz) in the southern Galactic plane. MALT-45 is to be conducted on the Australia Telescope Compact Array (ATCA), exploiting the newly available technologies made by the Compact Array Broadband Backend (CABB; \citealt{wilson11}). In particular, MALT-45 seeks to employ the CABB provided auto-correlations. Cross-correlation data are typically all that is used from an interferometer, as it provides the high resolution imagery not available from a single dish. However, drawbacks in interferometery include gaps in the observed uv-plane, meaning collected data is not completely represented. This is especially apparent for extended emission, such as \co, \nh and \cs, as these uv-plane gaps will not reveal the full extent of extended emission. Auto-correlation data alleviates this by combining each antenna in an interferometer as if they were all single dishes, leaving no gaps in the uv-plane. In this way, the CABB can be used to image extended emission with sensitivity similar to 6$\times$ Mopra (a single 22\,m dish radiotelescope similar to a single ATCA antenna). Hence, MALT-45 achieves the advantages of using an interferometer to accurately position interstellar masers associated with HMSF, while simultaneously mapping the Galaxy for extended emission.

MALT-45 is an untargeted Galactic plane survey, covering a total of 30 square-degrees, with a Galactic latitude width of $\pm0.5^{\circ}$. The survey will focus on the dense gas tracer \cs (1--0), and identify previously unknown sources of both Class I \choh and \sio (1--0) masers, which occur towards sites of star formation and evolved stars, respectively. The survey region of MALT-45 will be a subsection of HOPS (which covers a range $l=290^{\circ}$ through $l=0^{\circ}$ to $l=30^{\circ}$, $b=\pm0.5^{\circ}$). Other lines of possible interest are outlined in Table \ref{tab1}. The CABB allows a 64M-32k zoom mode configuration (see Section \ref{sec:pilotsurveyobs}), which will be used to provide fine velocity resolution, while also covering a large velocity range.

\subsection{Tracers}
The ground state transition ($J=1-0$) of \cs is a high density gas tracer ($4.6\times10^4$\,cm$^{-3}$; \citealt{evans99}) that can be used in conjunction with HOPS \nh data to identify the regions of high density in our Galaxy. Apart from the slightly higher critical density of \cs, compared to \nh, \cs is also known to freeze-out onto dust grains in the coldest and densest conditions \citep{tafalla02}, whereas \nh appears to be relatively robust against freeze-out \citep{bergin02}. \cs can be carefully compared to \nh to identify these cold and dense regions (low \cs/\nh ratio and \nh temperature) in our Galaxy, which identify the precursors to future star formation. We expect \cs to be optically thick at some positions, but we are fortunate that C$^{34}$S also occurs within the frequency band of MALT-45. A comparison of \cs and C$^{34}$S can give us a direct measure of the optical depth.

Methanol masers are separated into two classes: Class I and II. The better known and studied Class II masers are closely associated with HMSF \citep{walsh98}. The Class I masers are traditionally thought to be associated with outflows, offset from sites of star formation, although some evidence suggests there may be closer associations in some places \citep{voronkov06}. The strongest Class I methanol maser occurs at 44.069\,GHz, in the 7\,mm band, but it has not been searched using an untargeted survey. With the advent of 7\,mm receivers on the ATCA, we will compare their occurrence to other masers, such as the Class II masers from the Methanol Multibeam Survey (MMB; \citealt{green09}) and water masers from HOPS \citep{walsh11} in an untargeted way.

\sio masers are typically associated with evolved AGB stars, where they occur at a few stellar radii in the extended atmosphere \citep{elitzur92}. There are only three known sites of \sio masers towards star forming regions (Orion, W51 and Sgr B2), despite extensive searches \citep{zapata09}. However, \sio masers in Orion provide some evidence of a high mass accretion disk \citep{matthews07}. It is hoped that an untargeted survey for \sio masers will detect many previously unknown sources around late-type stars, and previously unknown masers associated with star formation. The ground state (thermal) \sio transition is commonly found in star formation regions and is a reliable tracer of outflows \citep{martin-pintado92}. MALT-45 will detect strong outflows from regions of HMSF, which will be the subject of future studies.

The CABB provides excellent sensitivity to continuum emission, even with very short integration times, given that two 2\,GHz bands are observed simultaneously. MALT-45 is in the unique position of being a high-frequency untargeted survey, simultaneously collecting auto-correlated data. Since continuum emission is typically extended, we aim to use the auto-correlation data to image the continuum with arcminute resolution. Additionally, the 7\,mm continuum cross-correlation data can be used to filter out extended emission to study ultracompact and hypercompact \hii regions. With the CABB, we will also be able to measure the spectral index of continuum sources between 42.2 and 49.2\,GHz. At MALT-45 frequencies, some continuum sources will be dominated by free-free emission and some by dust emission. The spectral index will allow us to discriminate between the two mechanisms of emission.

\subsection{The G333 GMC}
The MALT-45 pilot survey was conducted partially over the GMC associated with the \hii region RCW 106. RCW106 has a centre located roughly at $l=333^{\circ}$, $b=-0.5^{\circ}$. We are particularly interested in this GMC, commonly referred to as the G333 complex, located at a distance of 3.6\,kpc \citep{lockman79}. The G333 GMC has been thoroughly investigated for molecular spectral lines relevant to star formation \citep{bains06,wong08,lo09}, and has shown evidence of HMSF \citep{lo11}. See \citet{bains06}, \citet{wong08} and \citet{lo09} for a detailed description of the G333 complex. G333 is an ideal location to test MALT-45 before commencing the full survey, as we can expect extended \cs thermal emission and Class I \choh masers to be detected \citep{slysh94}.

\begin{table*}
  \begin{center}
  \caption{Spectral Lines in the 42.2 to 49.2\,GHz range. Radio recombination lines (RRL) are taken from \citet{lilley68}. All other rest frequencies are taken from the Cologne Database for Molecular Spectroscopy (CDMS; \citealt{muller05}).}
  \label{tab1}
  \begin{tabular}{ l c c c c }
    \hline
    Spectral Line   & Frequency & Pilot survey & Maser or & Beam Size \\
                    & (GHz)     & detection    & thermal? & (arcsec)\\
    \hline
    \sio (1--0) $v=3$            & 42.51934 & N & Maser & 66 \\
    \sio (1--0) $v=2$            & 42.82048 & Y & Maser & 66 \\
    H53\footnotesize{$\alpha$} (RRL) & 42.95197 & Y & Thermal & 65 \\
    \sio (1--0) $v=1$            & 43.12203 & Y & Maser & 65 \\
    \sio (1--0) $v=0$            & 43.42376 & Y & Thermal & 65 \\
    \choh 7(0,7)--6(1,6) A$^{+}$ & 44.06941 & Y & Maser (Class I) & 64 \\
    H51\footnotesize{$\alpha$} (RRL) & 48.15360 & N & Thermal & 58 \\
    C$^{34}$S (1--0)             & 48.20694 & Y & Thermal & 58 \\
    \choh 1$_1$--0$_0$ A$^{+}$   & 48.37246 & Y & Thermal & 58 \\
    \choh 1$_1$--0$_0$ E         & 48.37689 & N & Thermal & 58 \\
    OCS (4--3)                   & 48.65160 & N & Thermal & 58 \\
    \cs (1--0)                   & 48.99095 & Y & Thermal & 57 \\
    \hline
  \end{tabular}
  \end{center}
\end{table*}

\subsection{Pilot Survey Observations and Data Reduction}
\label{sec:pilotsurveyobs}
The first observations for MALT-45 were conducted on the ATCA over 36 hours from the 18th to 20th of March 2010, using the H168 array configuration. Additional MALT-45 observations were conducted on the 7th and 8th of June 2010, using a 6C array configuration. These observations were necessary to test the feasibility of MALT-45, as conducting an untargeted survey at this frequency will be demanding on time, and may yield poor data if not allowed to observe optimally.

A mosaic of source pointings was constructed to observe over one square-degree, with interleaved points spaced on a square grid of 42 arcsec, which is slightly larger than Nyquist sampled at the highest frequency (29 arcsec, assuming a beam of 57 arcsec). The mosaic contained roughly 12,000 points, and thus required short exposure times for each point to be completed within the allocated observation time, hence only 6 seconds of observation was allowed for each pointing. Prior testing to the pilot survey revealed that the ATCA required nominally 1.5 seconds settling time between adjacent points, thus leaving at most 4.5 seconds per pointing. This pilot survey was also necessary to test the 6 second cycle time used instead of the ATCA default 10 seconds.

Due to software limitations, a single mosaic of 12,000 discrete pointings was not possible on the ATCA. Instead, it was split into two half-mosaics, observed one after the other. However, an unfortunate miscalculation on the position of each half-mosaic caused an approximate half-degree gap between each mosaic. Nevertheless, an entire square-degree was mapped as planned. The centres of each half-mosaic are approximately $333.4^{\circ}$ and $334.1^{\circ}$ in Galactic longitude.

At the time of observation, the CABB was unable to provide the 64M-32k zoom windows, providing only broadband mode windows. The 64M-32k zoom windows, when completed and ready for use on the CABB, provide 16 simultaneous observing windows per 2048\,MHz broadband window, each with 2048 channels over 64\,MHz, yielding a 32\,kHz fine resolution. At the rest frequency of a Class I \choh maser, each channel of a zoom window provides a resolution of approximately 220\,\ms. Without the zoom modes, the ATCA provides two broadband windows simultaneously, each with 2048\,MHz of bandwidth, and 2048 channels. These windows were centred on frequencies of 43.2 and 48.2\,GHz, and were placed to observe \sio (1--0) masers, Class I \choh masers and \cs (1--0) thermal emission.  These spectral lines and hence the 2\,GHz band placements have been selected as being easiest to detect without sacrificing too many other lines of interest, and therefore some lines are neglected (eg. HCCCN at 45.49031\,GHz). As each channel corresponds to 1\,MHz in the spectrum, the approximate resolution at these frequencies is a coarse 6.8\,\kms. Future MALT-45 observations using the 64M-32k zoom windows on each of the 12 spectral lines listed in Table \ref{tab1} will provide much higher channel resolution, resulting in more detailed velocity information.

In this paper, we focus on the auto-correlation data rather than interferometric data, as this is the most innovative data processing aspect of the MALT-45 pilot survey. As auto-correlation data reduction is not common on the ATCA, the procedure to make this imagery possible was developed by the authors. These steps include:\\
\\
1. Using \textsc{MIRIAD} tasks \textsc{atlod} and \textsc{uvaver} to first select auto-correlations from the raw data, and select specific channel ranges of the data applicable to emission;\\
2. Converting extracted ATCA CABB autocorrelations from the \textsc{MIRIAD} dataset into an \textsc{sdfits} format as if these data resulted from a single-dish with a 6-beam receiver, one beam per ATCA antenna;\\
3. Removing the baseline of the single-dish data using \textsc{LiveData}. Each ``beam'' presented corresponds to an individual antenna from the ATCA;\\
4. Producing data cubes with \textsc{Gridzilla}.\\

\textsc{LiveData} parameters used include a referenced method with a mean estimator, a robust linear fit for baseline subtraction, and a Doppler frame of LSRK. The reference position used for baseline subtraction was G334.0$-$1.0, and was observed for 5 cycles (30 seconds) approximately every 34 minutes. \textsc{Gridzilla} parameters include a beam FWHM of 1 arcmin, a kernel FWHM of 1.5 arcmin, a cutoff radius of 0.8 arcmin and a mean gridding statistic.

Once cubes are produced, they may be further manipulated in \textsc{MIRIAD} for analysis, such as producing peak-intensity maps. Cubes needed to be produced individually for each emission line, as the channel range that could be processed was small. At the time of observation, the CABB also had bad channel issues with the broadband windows, specifically, on the channel range 250-500, and all multiples of 512.

It was found that data from antenna 6 of the ATCA (CA06) had increased noise levels by a factor of 3.5. The removal of CA06 data from the other antenna data improved the quality of images substantially. The source of the poorer performance of CA06 is suspected to be due to the different dish surface, and its poor reception to 7\,mm emission.

\textsc{MIRIAD}\footnote{http://www.atnf.csiro.au/computing/software/miriad/}, \textsc{LiveData} and \textsc{Gridzilla}\footnote{http://www.atnf.csiro.au/computing/software/livedata/index.html} are software packages managed and maintained by CSIRO Astronomy and Space Science.

\section{Results}

\begin{figure*}
  \includegraphics[width=0.8\textwidth]{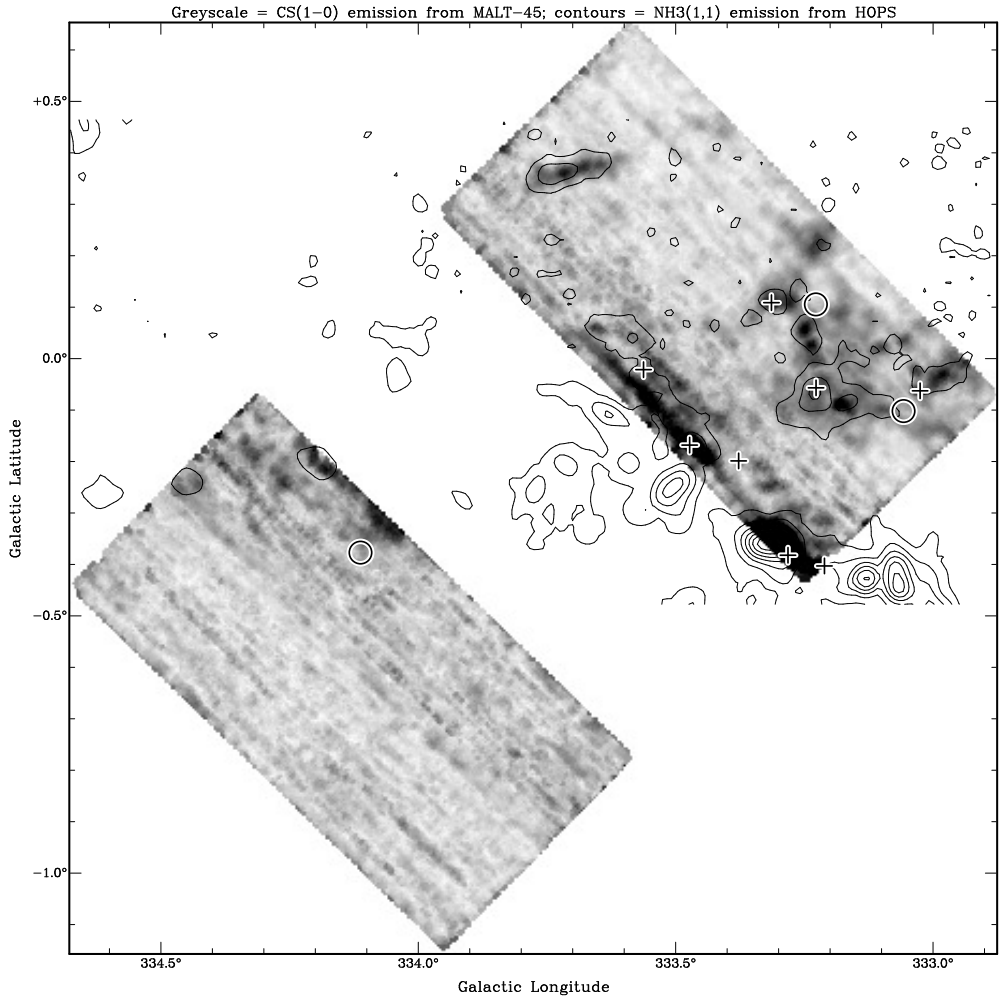}
  \caption{Auto-correlated \cs peak-intensity map, overlaid with HOPS thermal \nh (1,1) contours. Contour levels are 10\%, 20\%, ...90\% of 1.09\,K in units of antenna temperature. Circles represent \sio $v=2$ maser positions, while plus symbols ($+$) designate Class I \choh maser positions.}
  \label{fig:cs}
\end{figure*}

\begin{figure*}
  \includegraphics[width=0.8\textwidth]{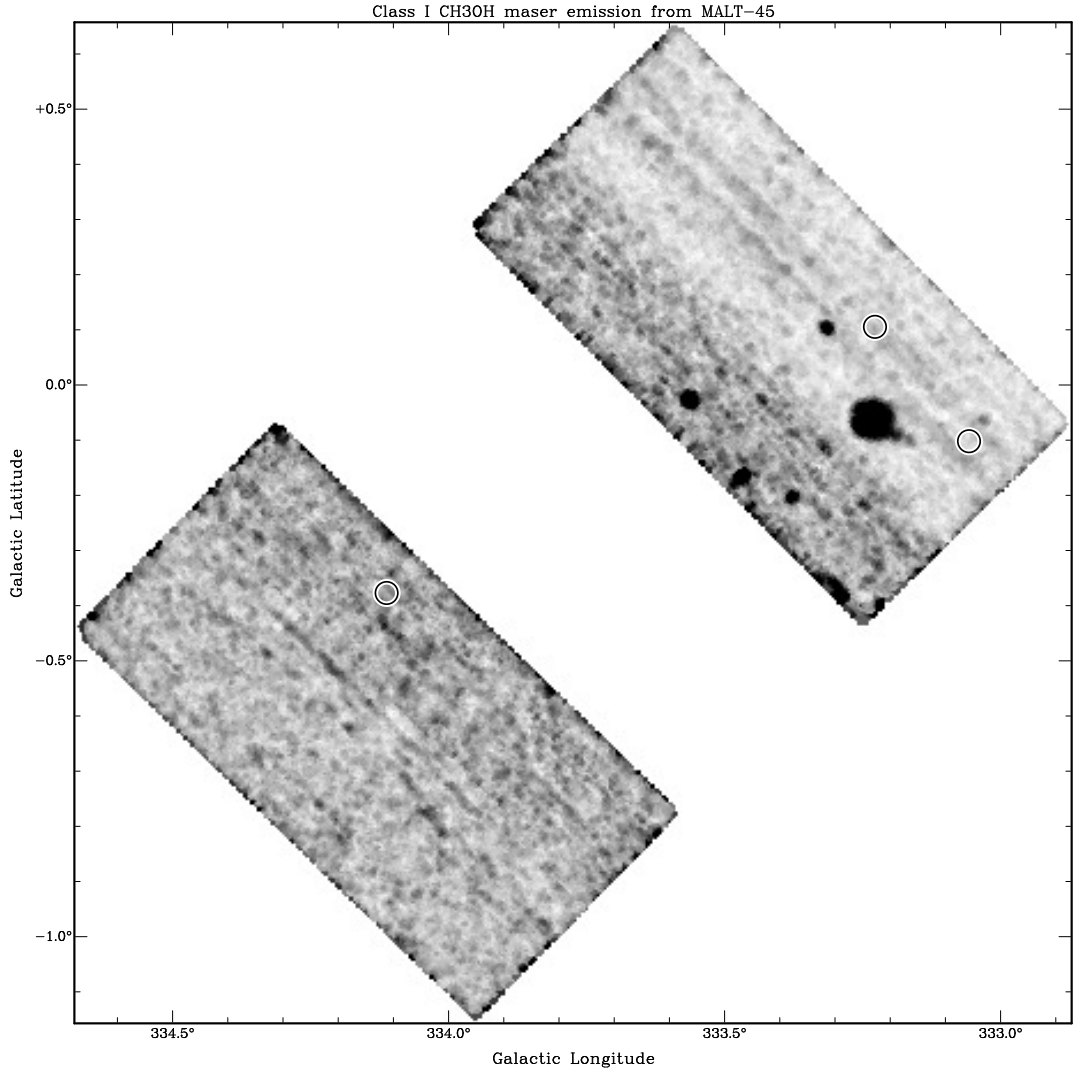}
  \caption{Auto-correlated Class I \choh maser peak-intensity map. Circles represent \sio $v=2$ maser positions.}
  \label{fig:choh}
\end{figure*}

\begin{figure*}
  \includegraphics[width=\textwidth]{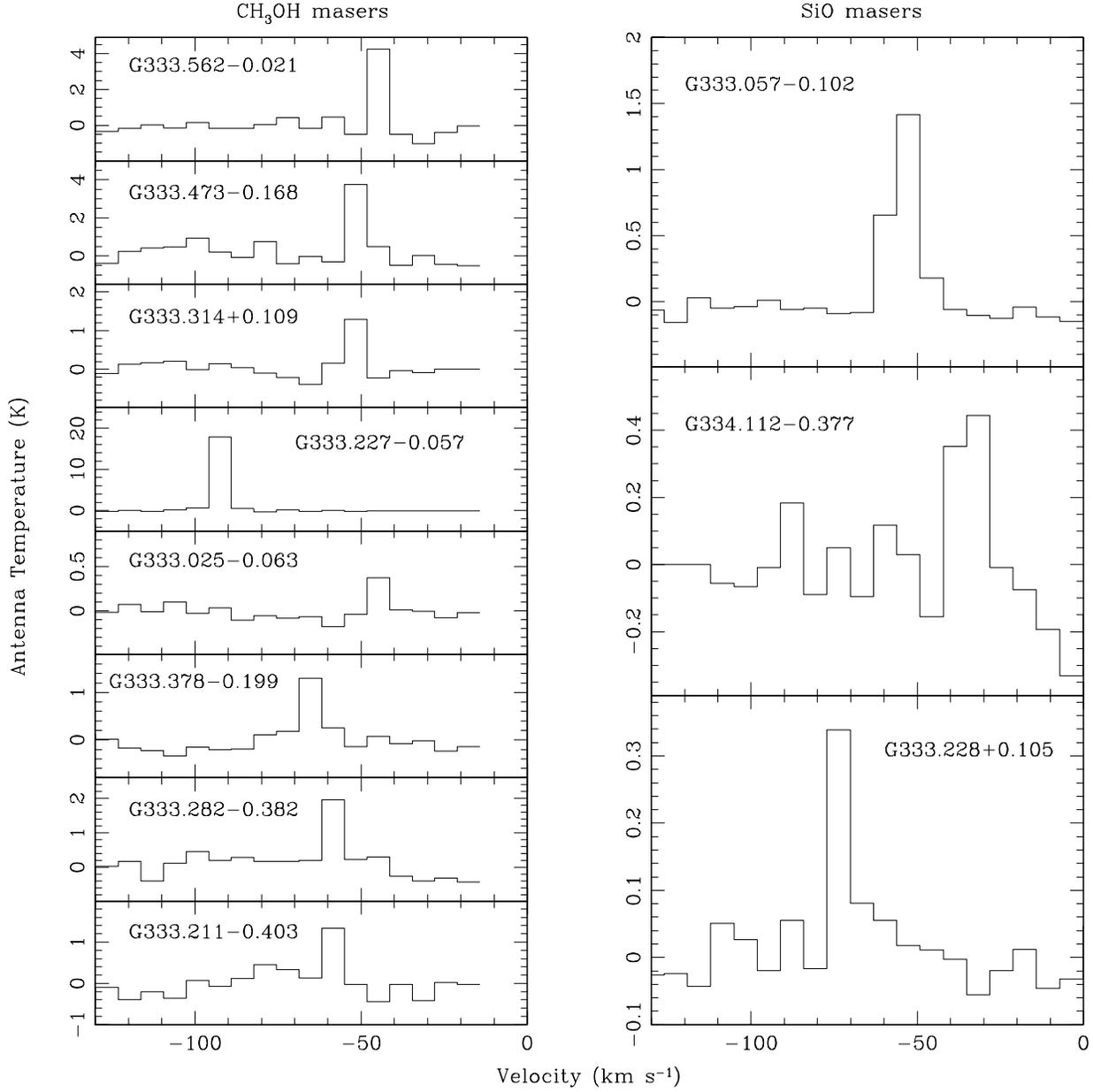}
  \caption{Auto-correlated spectra for each Class I \choh (left) and \sio $v=2$ maser (right) detected. The deviations in the baselines are a result of poor weather. The channel width is approximately 6.8\,\kms.}
  \label{fig:spectra}
\end{figure*}

\begin{figure*}
  \includegraphics[width=0.8\textwidth]{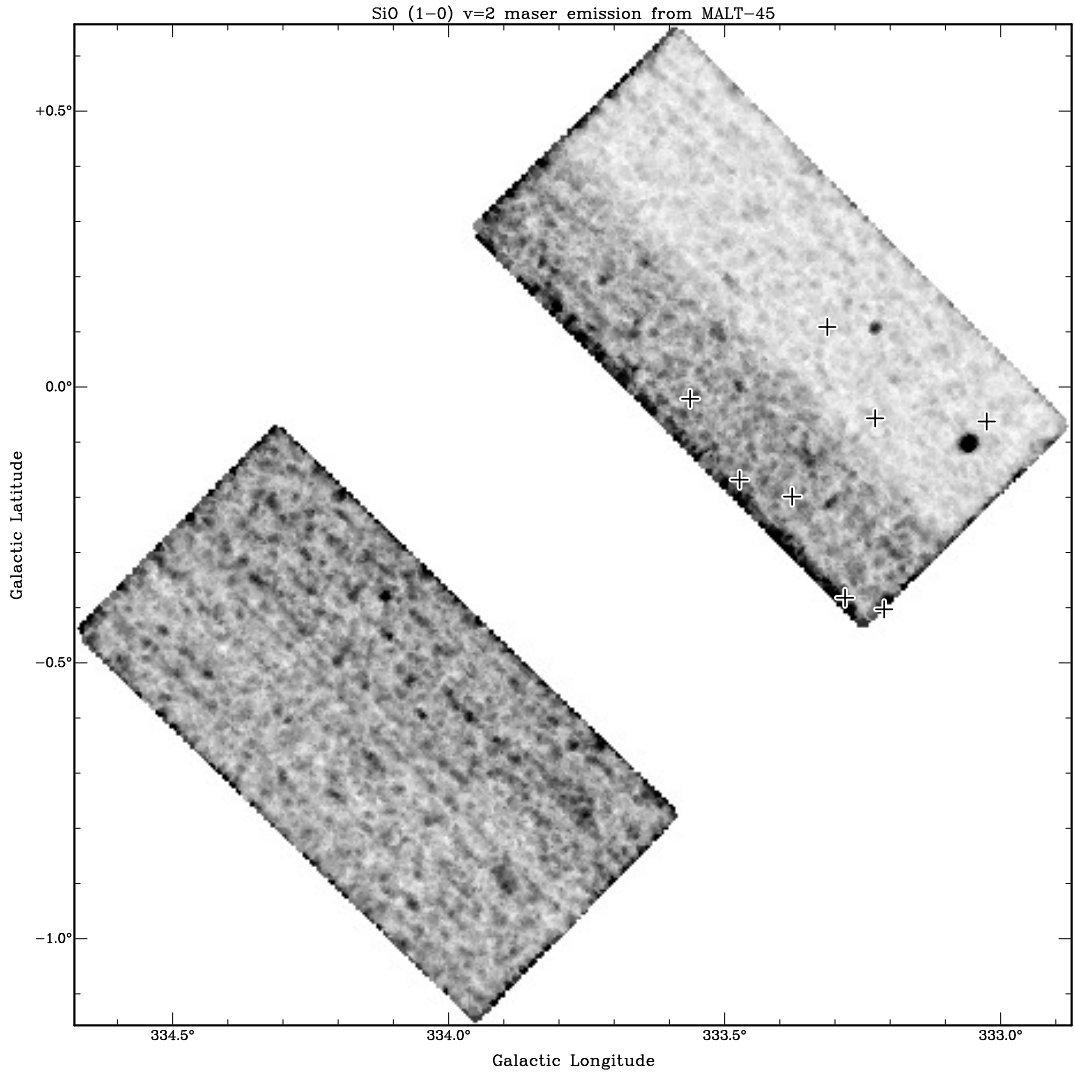}
  \caption{Auto-correlated \sio (1--0) $v=2$ maser peak-intensity map. Plus symbols ($+$) represent Class I \choh maser positions.}
% The \sio masers are located at (333.057, -0.102), (333.228, +0.105) and (334.112, -0.377).
  \label{fig:sio}
\end{figure*}

\begin{figure*}
  \includegraphics[width=\textwidth]{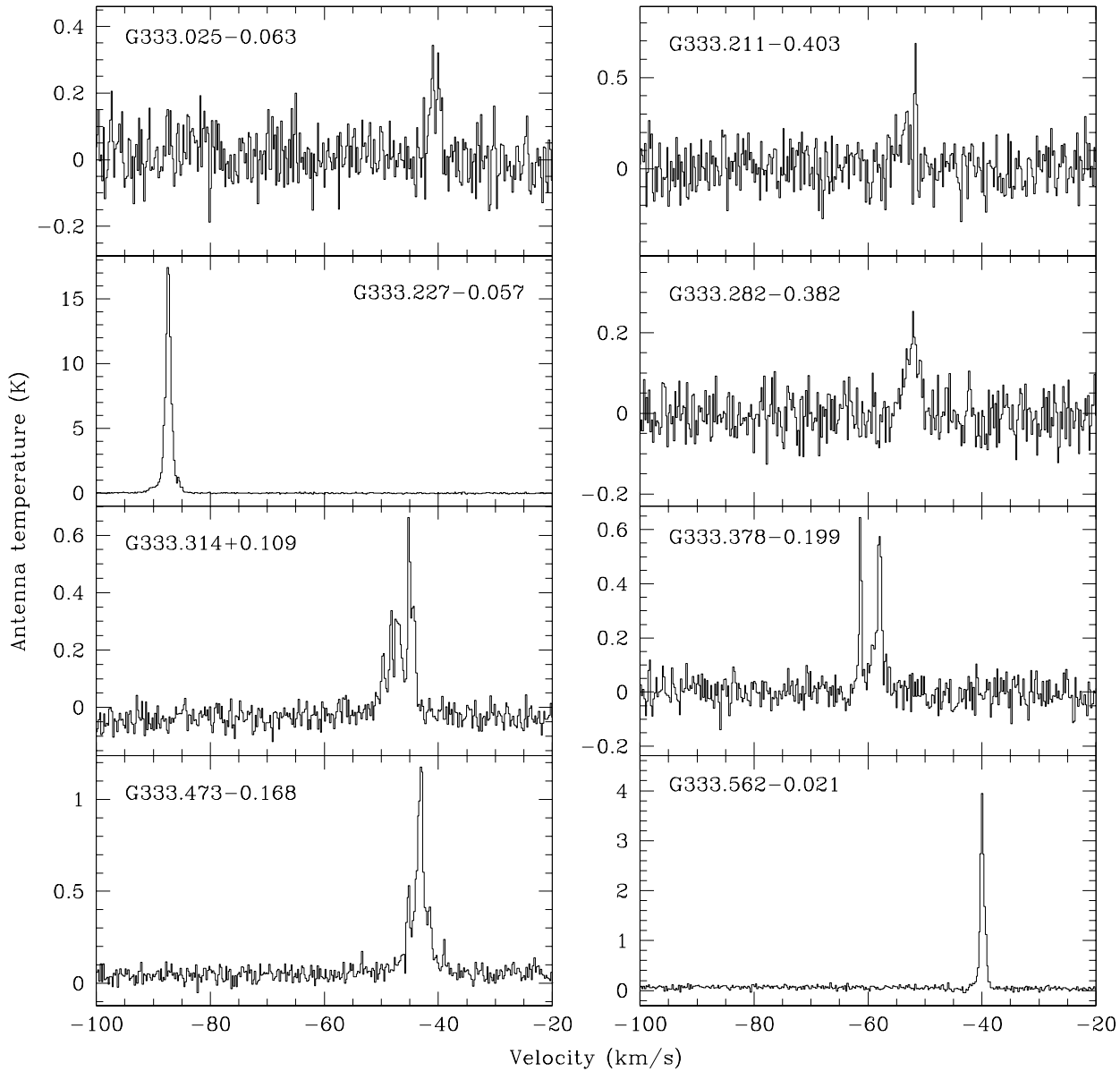}
  \caption{Follow-up Class I \choh maser spectra from the Mopra radiotelescope.}
  \label{fig:mopra_choh}
\end{figure*}

\begin{figure*}
  \includegraphics[width=0.9\textwidth]{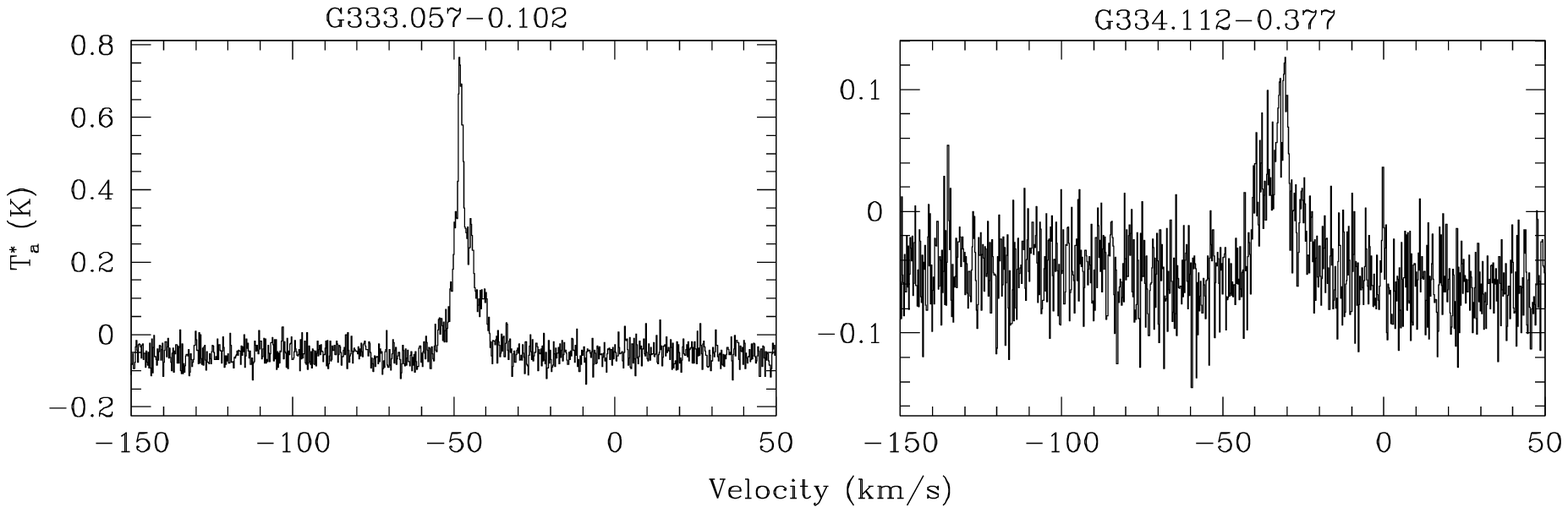}
  \caption{Follow-up \sio maser spectra from the Mopra radiotelescope.}
  \label{fig:mopra_sio}
\end{figure*}

\begin{table*}
  \begin{center}
    \caption{Properties of detected MALT-45 maser emission as determined by Mopra.}
    \label{tab2}
    \begin{tabular}{ c c c c c c c l }
      \hline
      Source Name & RA      & Dec     & Peak Antenna & Peak & \multicolumn{2}{c}{Velocity Range} & Maser Detected \\
                  & (J2000) & (J2000) & Temperature  & Velocity & Min & Max       & \\
                  & (h m s) & ($^\circ$ $^\prime$ $^{\prime\prime}$) & (K) & (\kms) & \multicolumn{2}{c}{(\kms)} & \\
      \hline
      G333.025$-$0.063 & 16 18 55.6 & -50 24 03.6 & 0.34 & $-$41 & $-$42 & $-$39 & Class I \choh \\
      G333.057$-$0.102 & 16 19 15.0 & -50 24 12.1 & 0.84 & $-$48 & $-$45 & $-$50 & \sio $v=1,2$ \\
      G333.211$-$0.403 & 16 21 15.6 & -50 30 45.1 & 0.69 & $-$52 & $-$54 & $-$51 & Class I \choh \\
      G333.227$-$0.057 & 16 19 48.6 & -50 15 17.2 & 18 & $-$87 & $-$86 & $-$88 & Class I \choh \\
      G333.282$-$0.382 & 16 21 29.0 & -50 26 52.0 & 0.25 & $-$52 & $-$55 & $-$50 & Class I \choh \\
      G333.314$+$0.109 & 16 19 28.8 & -50 04 45.0 & 0.70 & $-$45 & $-$45 & $-$48 & Class I \choh \\
      G333.378$-$0.199 & 16 21 05.8 & -50 15 00.9 & 0.65 & $-$61 & $-$62 & $-$56 & Class I \choh \\
      G333.473$-$0.168 & 16 21 22.7 & -50 09 42.6 & 1.1 & $-$43 & $-$42 & $-$44 & Class I \choh \\
      G333.562$-$0.021 & 16 21 08.1 & -49 59 47.8 & 3.8  & $-$40 & $-$39 & $-$41 & Class I \choh \\
      G334.112$-$0.377 & 16 25 06.7 & -49 51 28.5 & 0.14 & $-$30 & $-$29 & $-$34 & \sio $v=1,2$ \\
      \hline
    \end{tabular}
    \medskip \\
  \end{center}
\end{table*}

\begin{figure*}
  \includegraphics[bb=0 0 500 349]{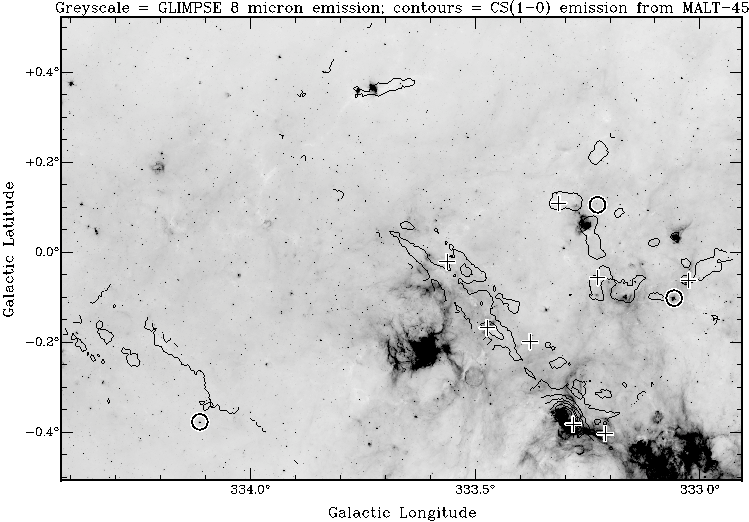}
  \caption{\emph{Spitzer} GLIMPSE 8.0$\mu$m greyscale image with MALT-45 pilot survey \cs (1--0) thermal emission in black contours, and positions of detected \choh (+) and \sio (circles) masers. Contour levels are 7\%, 17\%, ...87\% of 0.72\,K, in units of antenna temperature.}
  \label{fig:glimpse}
\end{figure*}

\begin{figure*}
  \subfigure{
    \includegraphics[width=0.47\textwidth]{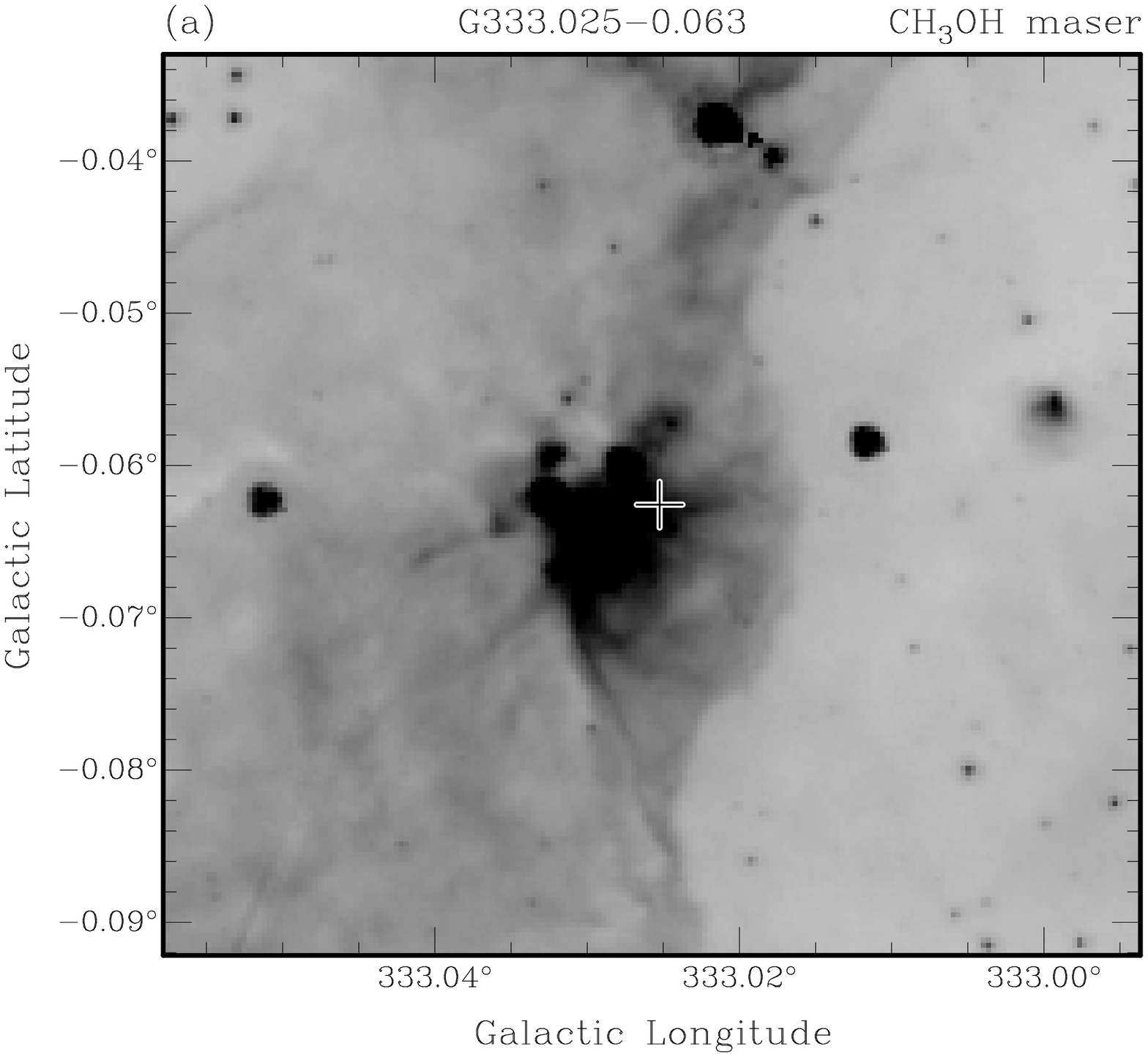}
  }
  \subfigure{
    \includegraphics[width=0.47\textwidth]{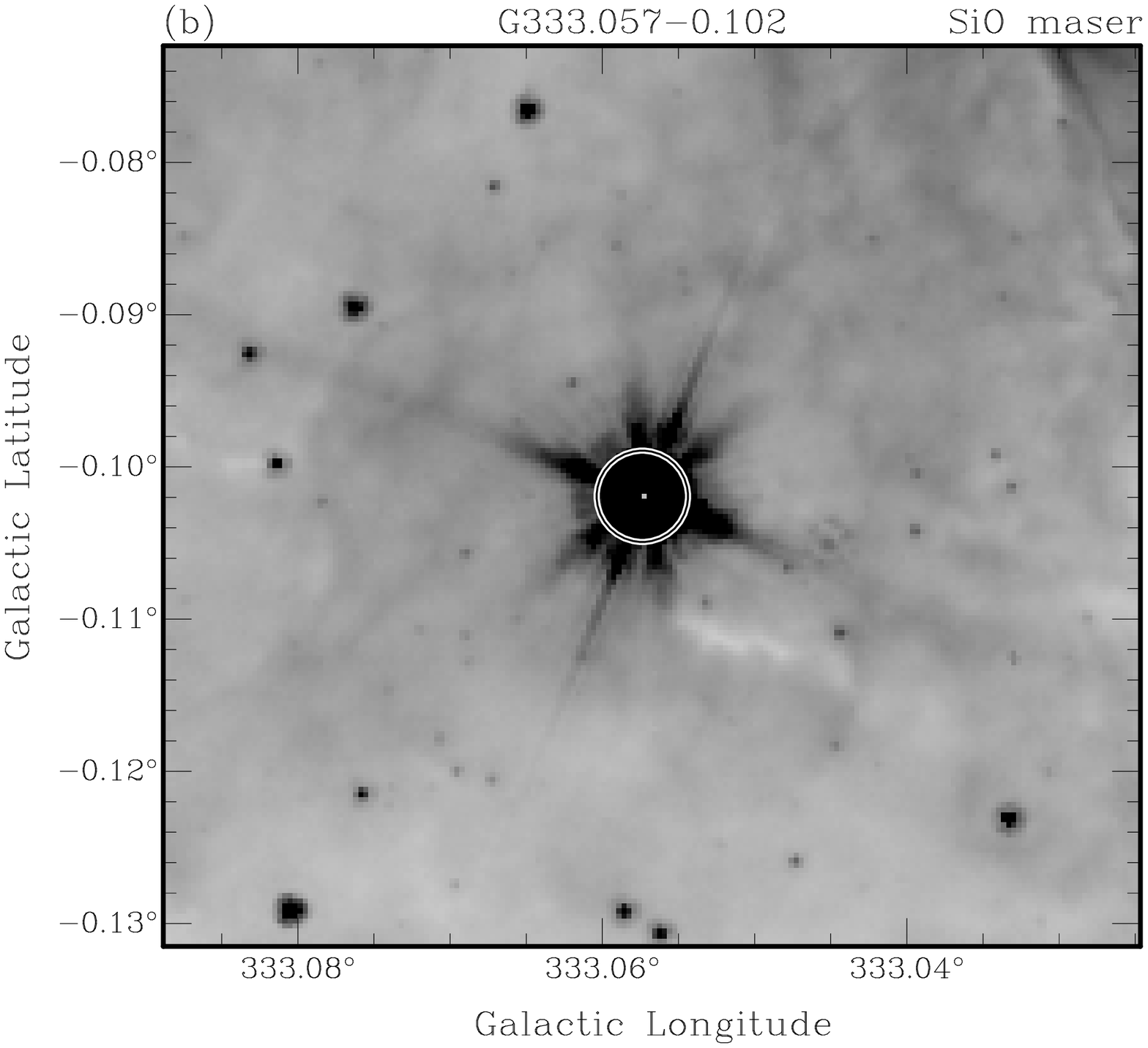}
  }
  \subfigure{
    \includegraphics[width=0.47\textwidth]{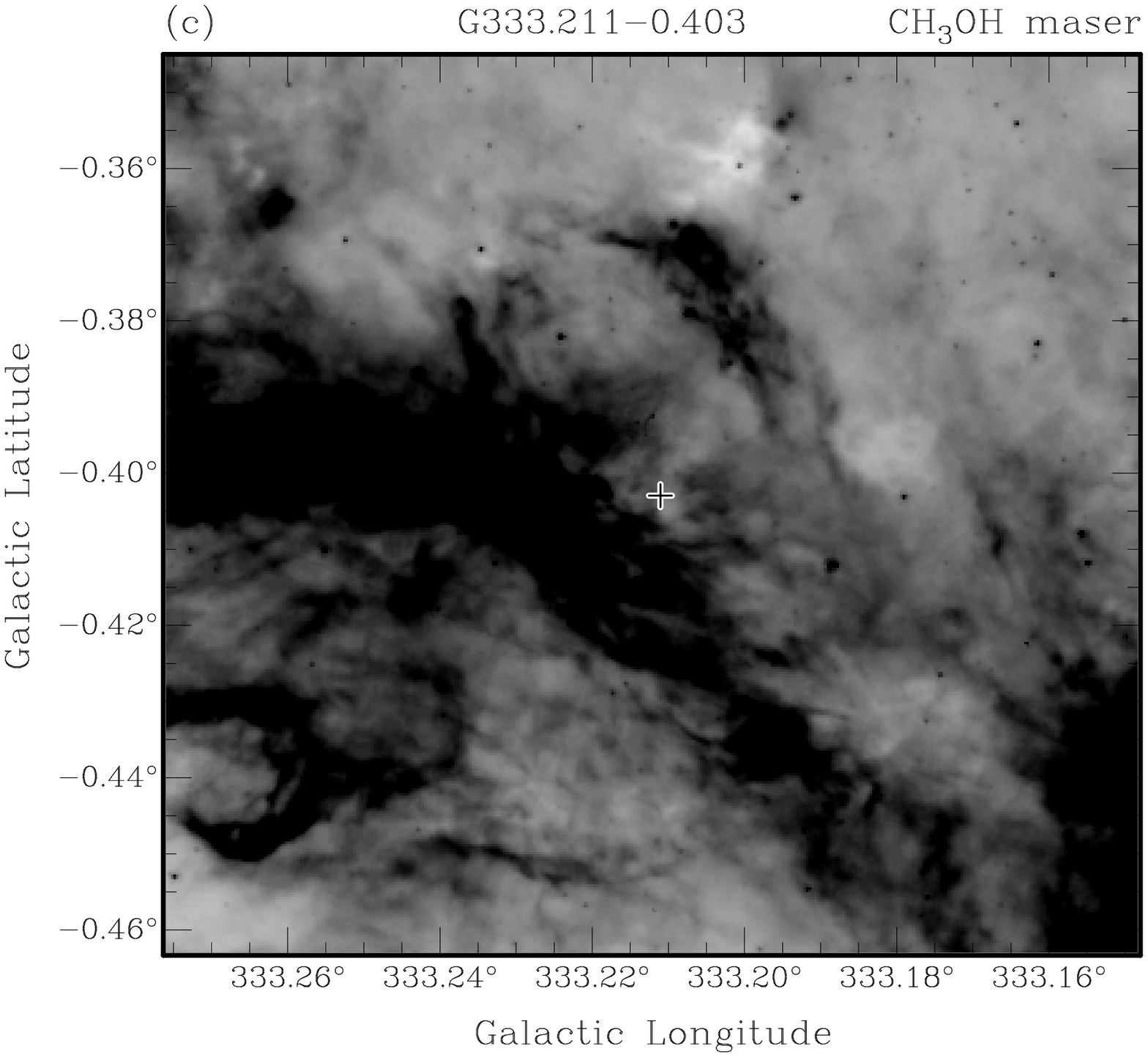}
  }
  \subfigure{
    \includegraphics[width=0.47\textwidth]{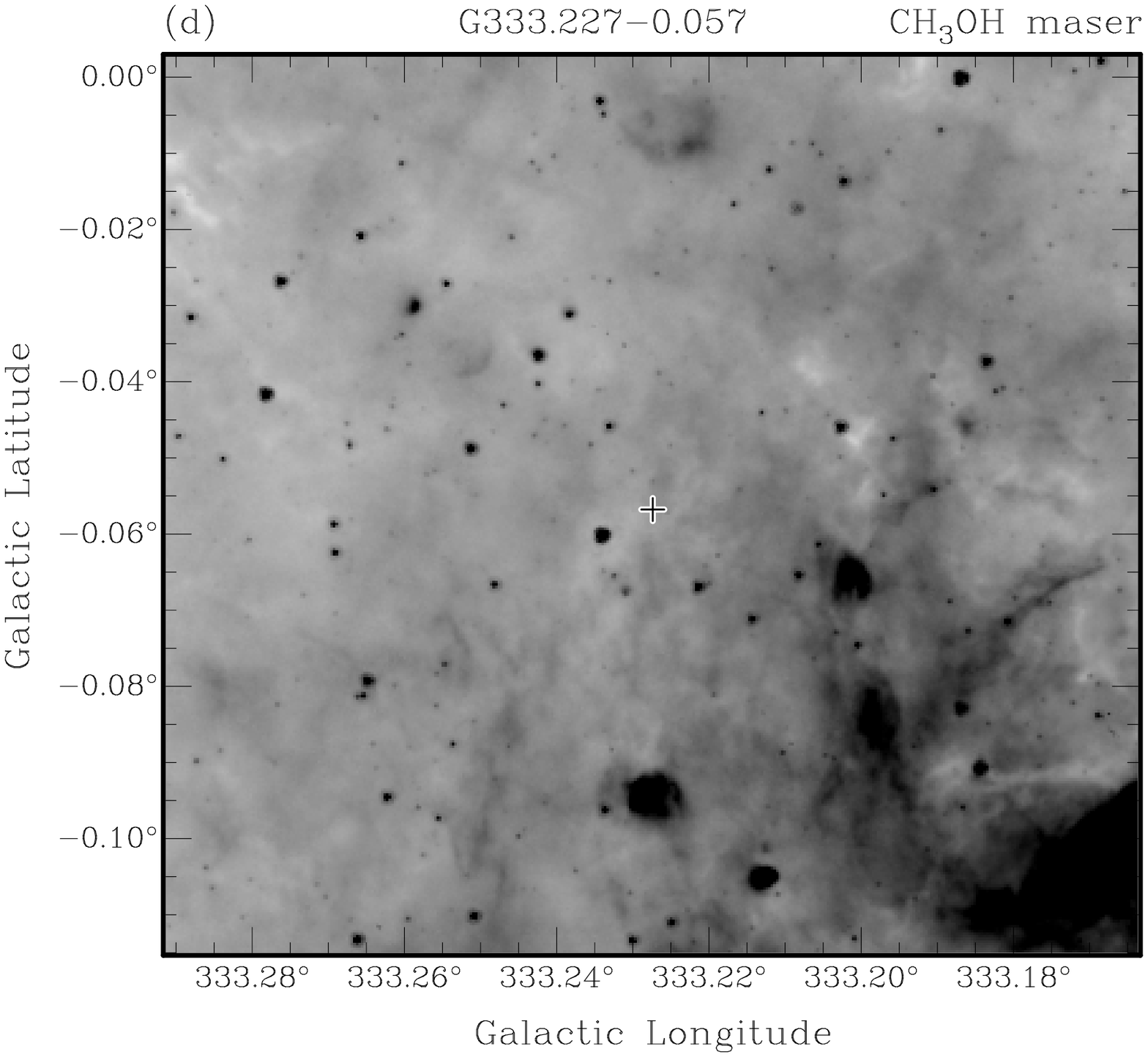}
  }
  \subfigure{
    \includegraphics[width=0.47\textwidth]{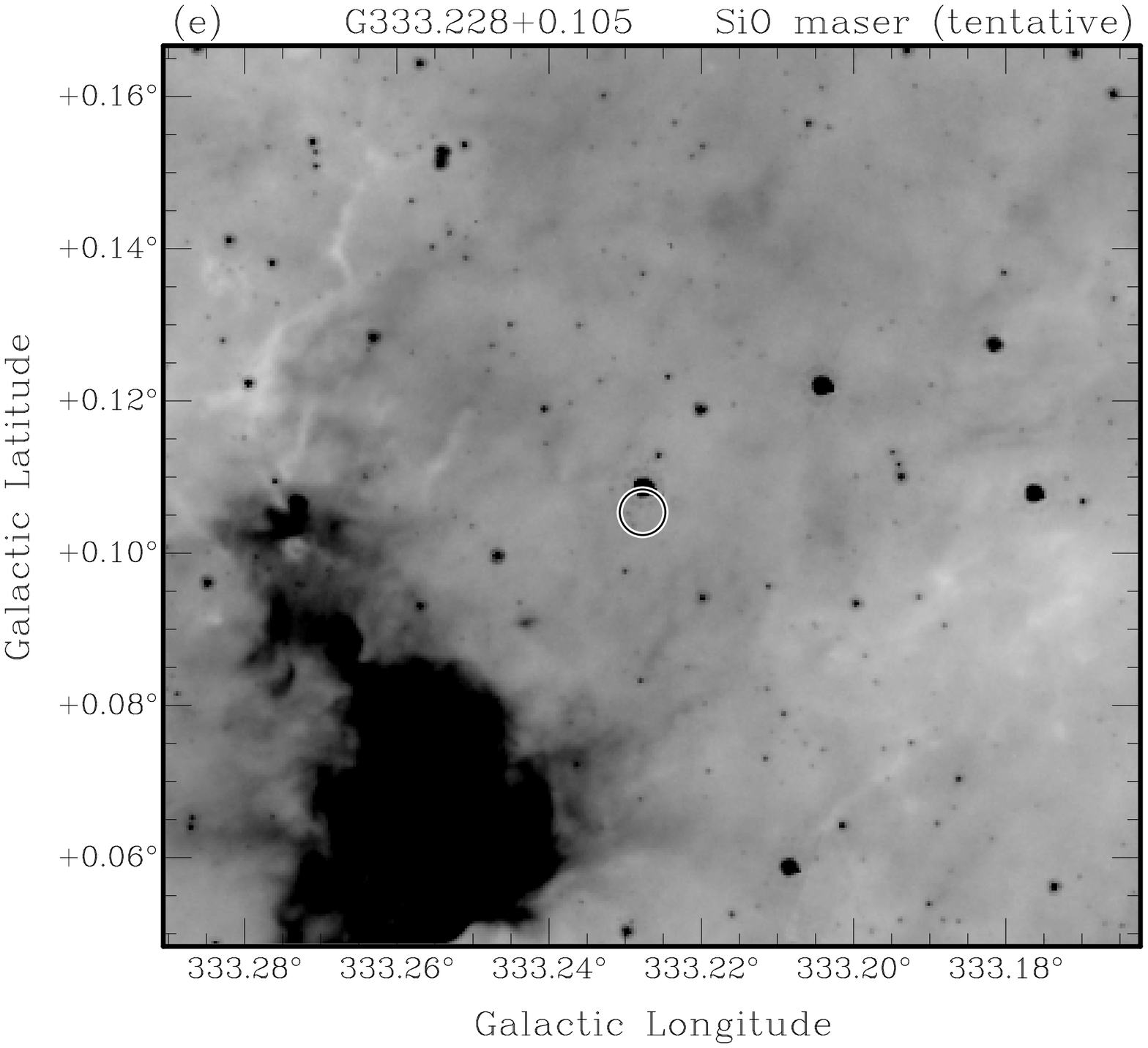}
  }
  \subfigure{
    \includegraphics[width=0.47\textwidth]{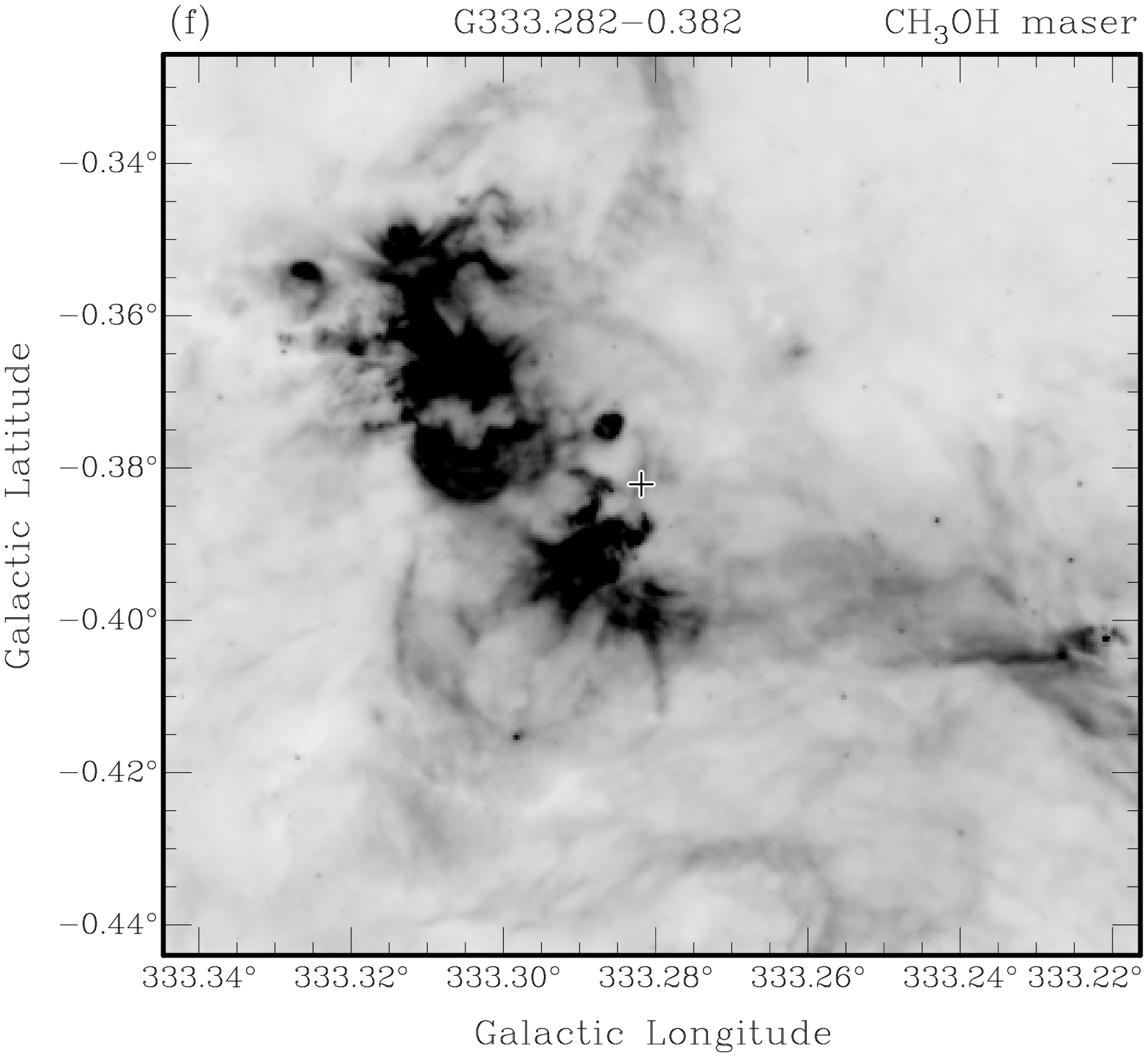}
  }
  \caption{GLIMPSE 8\,$\mu$m images of each region of maser detection. Plus ($+$) symbols designate positions of Class I \choh masers, and circles represent \sio maser positions.}
  \label{fig:closeglimpse}
\end{figure*}

\begin{figure*}
  \subfigure{
    \includegraphics[width=0.47\textwidth]{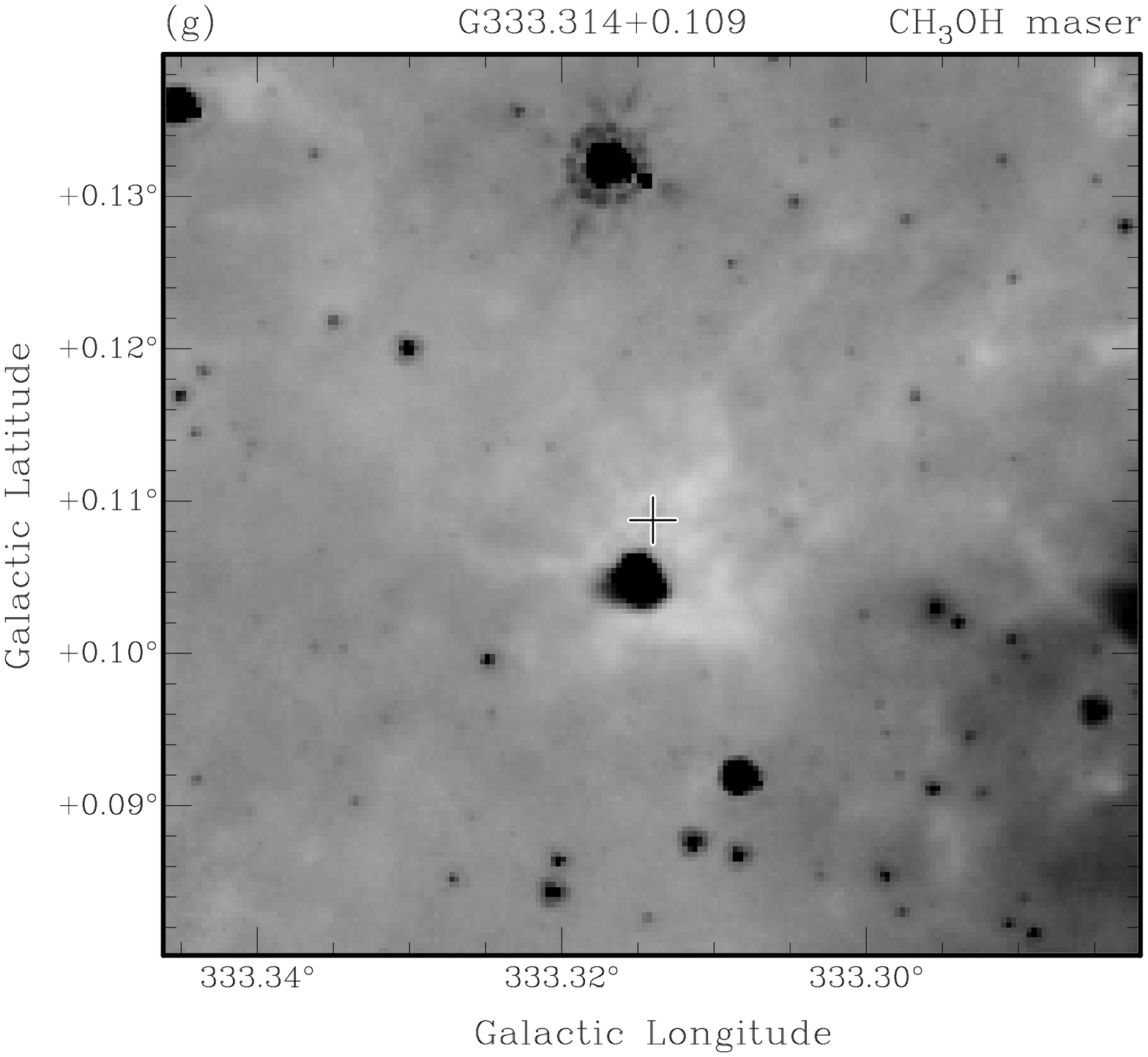}
  }
  \subfigure{
    \includegraphics[width=0.47\textwidth]{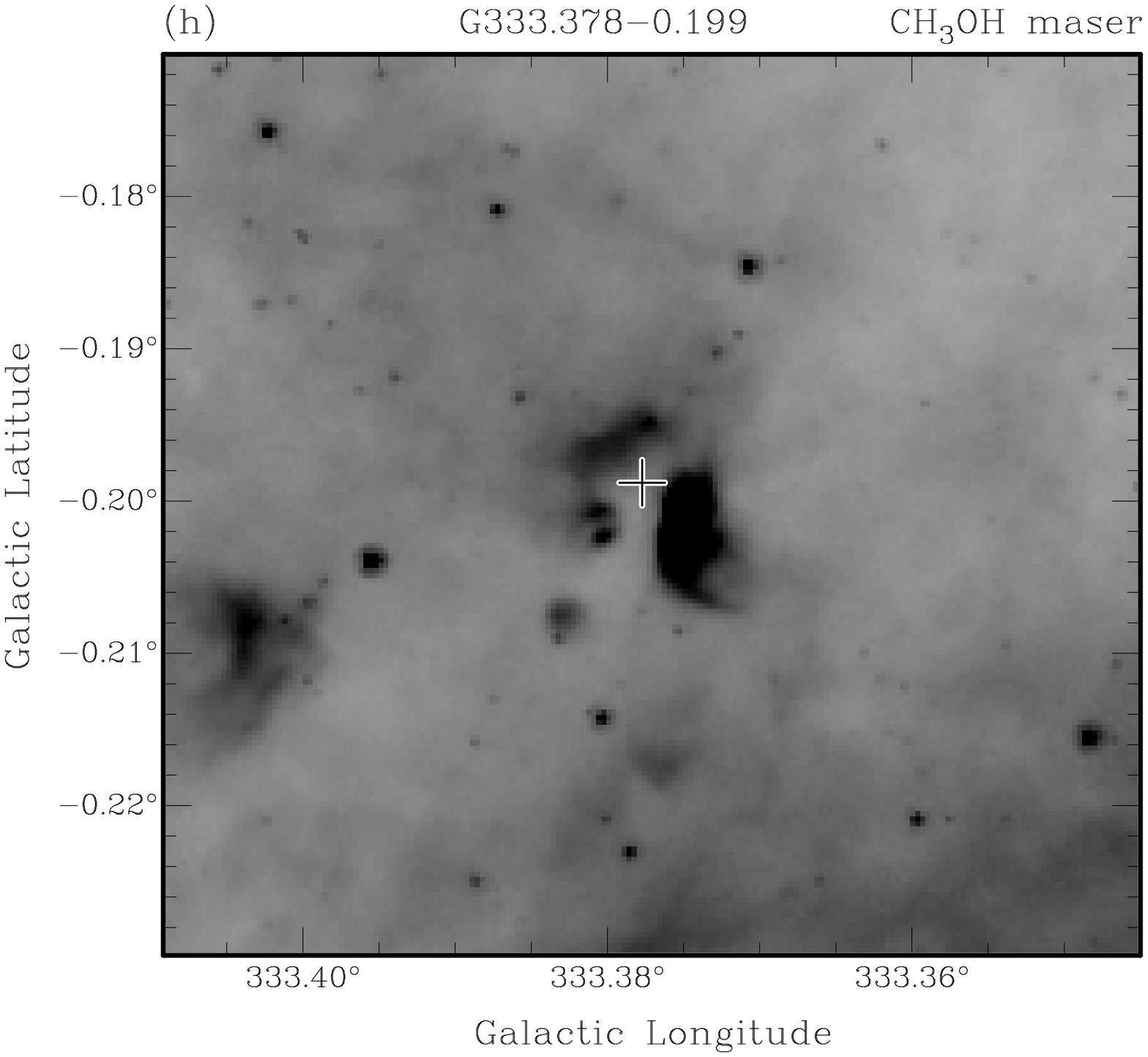}
  }
  \subfigure{
    \includegraphics[width=0.47\textwidth]{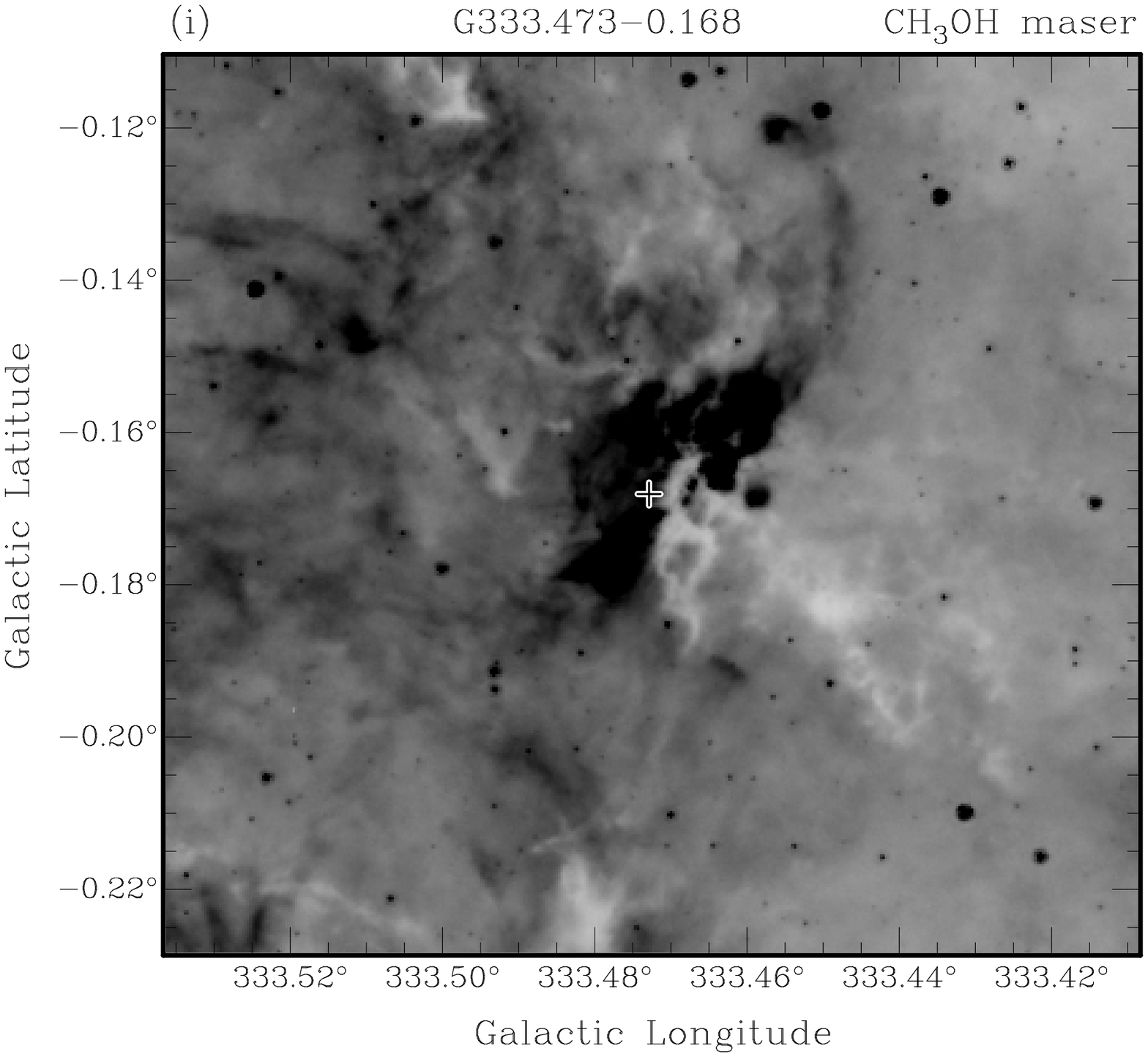}
  }
  \subfigure{
    \includegraphics[width=0.47\textwidth]{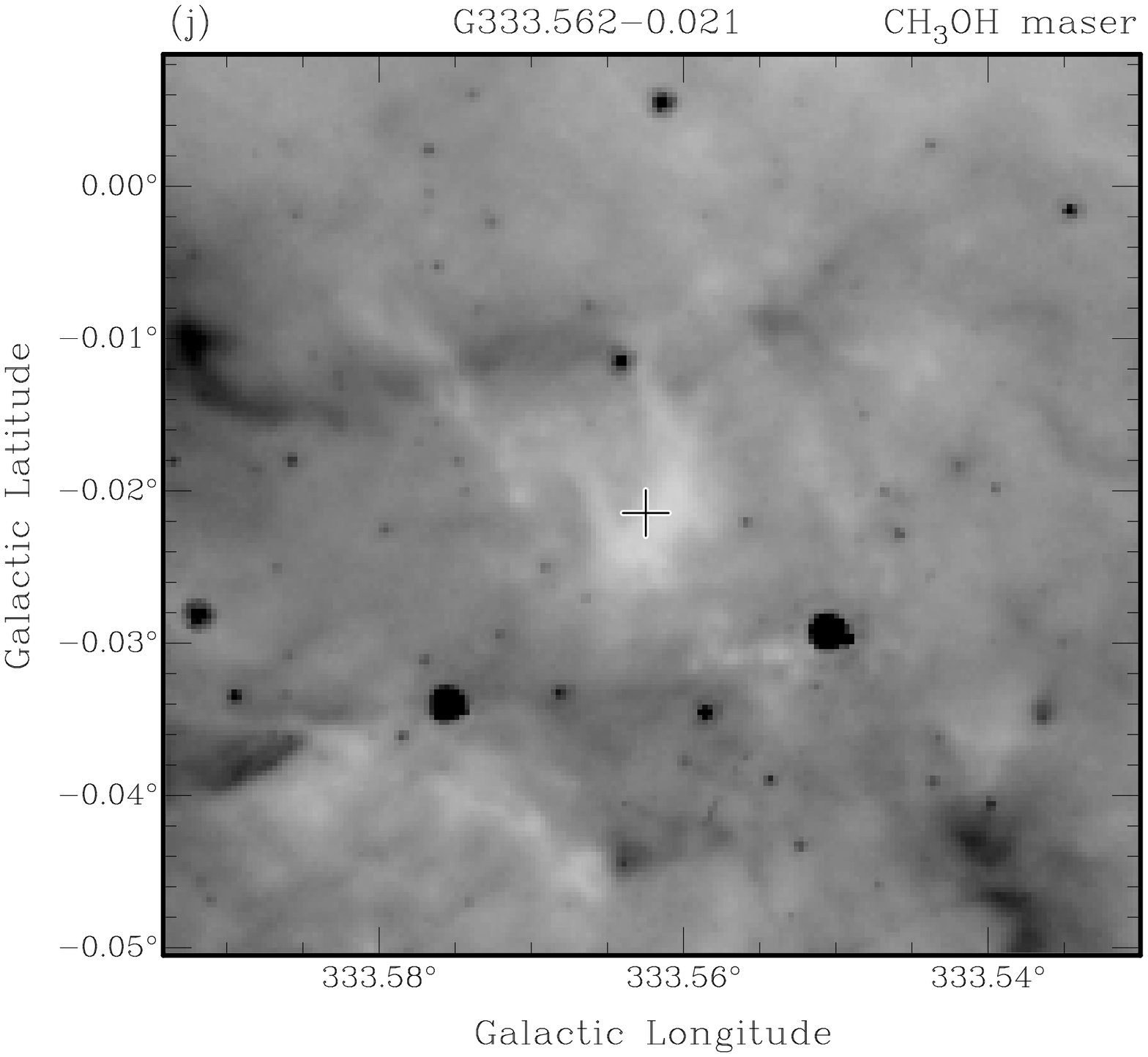}
  }
  \subfigure{
    \includegraphics[width=0.47\textwidth]{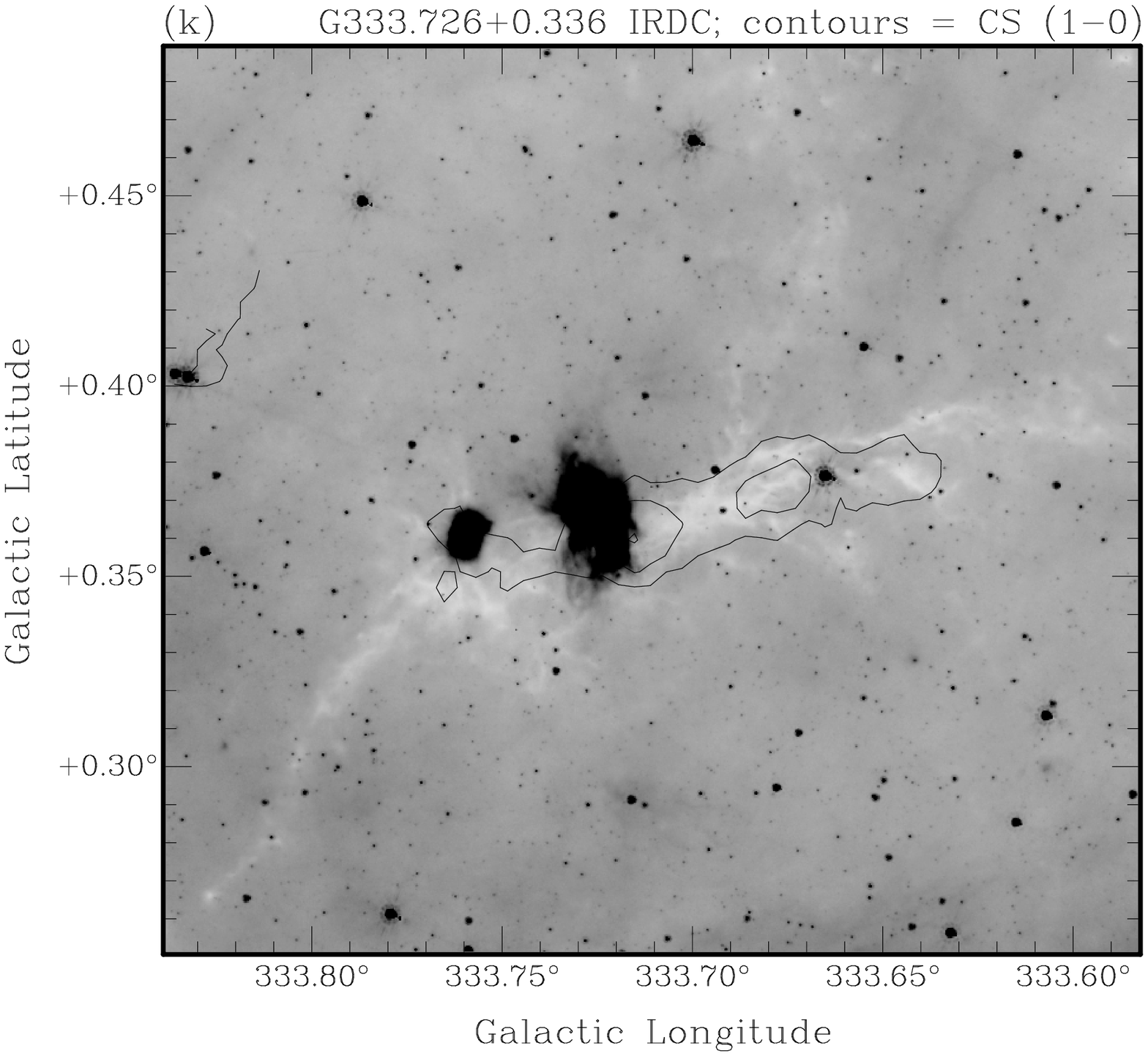}
  }
  \subfigure{
    \includegraphics[width=0.47\textwidth]{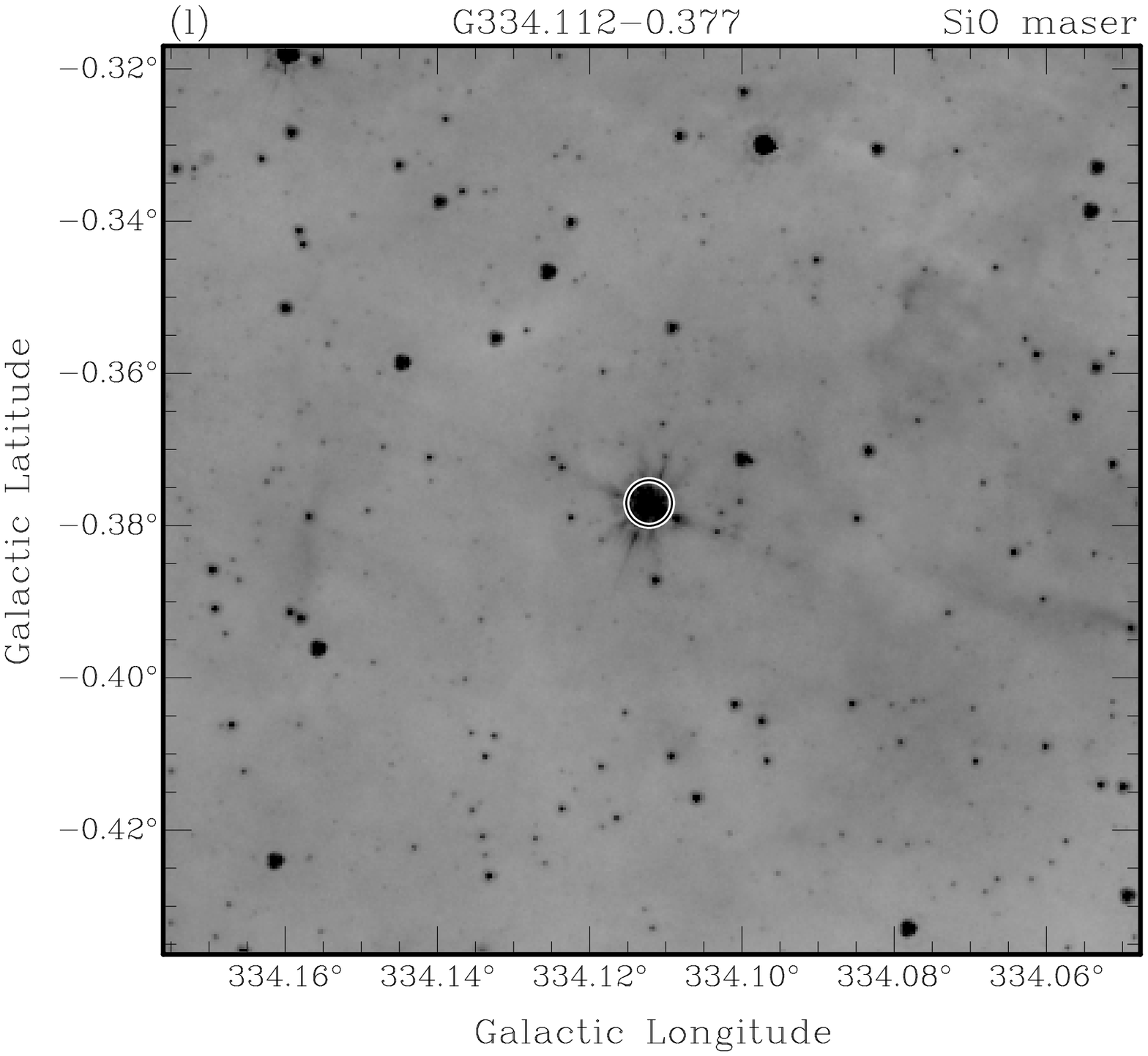}
  }
  \contcaption{(k) contours are \cs (1--0) emission at 7, 9 and 11\% of 0.72\,K, in units of antenna temperature.}
\end{figure*}

Auto-correlated data cubes were produced for each of the lines listed in Table \ref{tab1} from the ATCA broadband data. We detected emission in \cs (1--0) (Figure \ref{fig:cs}), Class I \choh masers at 44\,GHz (Figures \ref{fig:choh} and \ref{fig:spectra}), vibrationally excited \sio masers (Figures \ref{fig:spectra} and \ref{fig:sio}), as well as occasional detections of H53{\footnotesize{$\alpha$}}, \sio (1--0) $v=0$ and \choh 1$_1$--0$_0$ A$^{+}$. Radio recombination lines H53{\footnotesize{$\alpha$}} and H51{\footnotesize{$\alpha$}} were anticipated detections, as the brightest H69{\footnotesize{$\alpha$}} detection in the Galactic plane found by HOPS is located at (333.609, -0.211). The survey region of this pilot survey did not cover the known regions of emission, but some H53{\footnotesize{$\alpha$}} recombination line emission was found cospatial with the Class I \choh maser located at (333.281, -0.382).

As only the broadband modes were available at the time of observation, the results provide a coarse spectral resolution of about 6.8\,\kms. Fortunately, some observing time was permitted on the Mopra radiotelescope to briefly follow-up some of the detected Class I \choh and \sio masers. Observations on Mopra were carried out using a standard position-switch mode, where each on-source position was observed for 5 minutes, resulting in a higher sensitivity than the ATCA observations. Maser emission characteristics are detailed in Table \ref{tab2} and Figures \ref{fig:mopra_choh} and \ref{fig:mopra_sio}.

The RMS noise level was calculated, based on the 5 antennae used in auto-correlated data, to be about 30\,mJy\,beam$^{-1}$ (2\,mK\,beam$^{-1}$ on the main beam brightness temperature scale) per single 6.8\,\kms channel.

By comparing the positions of the same maser feature in G333.227-0.057, the estimated astrometric errors are derived from new ATCA data from Voronkov et al. (2012, in preparation). We find an offset of 9 arcseconds and so assign the estimated astrometric error of the pilot survey work to be 10 arcseconds.

\subsection{\cs (1--0) at 48.990\,GHz}
The main isotopologue of \cs was detected in extended emission across a large portion of the survey region (Figure \ref{fig:cs}). Also shown in Figure \ref{fig:cs} is \nh (1,1) emission, as detected in HOPS \citep{purcell12}. The \cs emission appears to closely follow the \nh emission, as might be expected from the two quiescent dense gas tracers (eg. \citealt{purcell09}). The strongest \cs emission occurs cospatial with the G333 GMC, which is close to the edge of the surveyed area around (333.3, $-$0.4).

Extended \cs emission is seen between Galactic longitudes of 329.9$^{\circ}$ and 333.3$^{\circ}$ and Galactic latitudes $-$0.2$^{\circ}$ and $+$0.3$^{\circ}$. Within this region \nh emission is detected, but covers a much smaller area than the \cs emission, likely attributable to the relatively poor sensitivity of HOPS.

At (333.70, $+$0.35), we see a filamentary \cs structure that is oriented close to the Galactic plane. This structure is also seen in \nh emission (Figure \ref{fig:cs}), as well as an infrared dark cloud (Figure \ref{fig:glimpse}). See section \ref{sec:comparison} for more details.

\subsection{Class I \choh 7(0,7)--6(1,6) A\mbox{\boldmath{$^{+}$}} masers at 44.069\,GHz}
In the survey region, we found eight possible Class I \choh masers, whose properties are listed in Table \ref{tab2} and shown as greyscale emission in Figure \ref{fig:choh}, and positions are shown in Figures \ref{fig:cs}, \ref{fig:sio}, \ref{fig:glimpse} and \ref{fig:closeglimpse} as plus ($+$) symbols. Most of the masers (5 of 8) are cospatial with the G333 GMC and have line of sight velocities within 10\,\kms of the GMC systemic velocity. They are thus likely to be associated with the star formation occurring in and around the GMC. The other three appear cospatial with extended \cs emission. The CABB broadband spectra for each maser are provided in Figure \ref{fig:spectra}. For each of these detections, we managed to acquire time on the Mopra radiotelescope to re-observe the \choh spectra, as shown in Figure \ref{fig:mopra_choh}. With the superior spectral resolution of Mopra, more convincing detections are made and finer velocity detail becomes apparent. Of all the \choh detections, G333.227$-$0.057 has been previously reported as a detection at 44\,GHz by \citet{slysh94} and G333.473$-$0.168 by \citet{voronkov12}.

\subsection{\sio (1--0) \mbox{\boldmath{$v=1,2$}} masers at 43.122 and 42.820\,GHz}
\sio emission in the vibrationally excited transitions, which are known to show maser activity, was tentatively detected towards three positions, shown as circles in Figures \ref{fig:cs}, \ref{fig:choh}, \ref{fig:glimpse}, and \ref{fig:closeglimpse}, and as greyscale in Figure \ref{fig:sio}. All three positions appear to exhibit emission in the $v=2$ transition and two of the three in the $v=1$ transition (the exception being G333.228$+$0.105). Spectra for the $v=2$ transition are shown in Figure \ref{fig:spectra}. Again, emission is restricted to a small number of channels, using the CABB broadband mode. Each of the three tentative masers were re-observed with Mopra and high-resolution spectra are provided for two in Figure \ref{fig:mopra_sio}, where more detail of the spectra can be seen. The maser at G333.228$+$0.105 was not detected by Mopra, leaving its detection status by the CABB as tentative. Figures \ref{fig:glimpse} and \ref{fig:closeglimpse} show the locations of the \sio masers with respect to infrared emission from GLIMPSE. Each \sio maser is cospatial with a bright infrared star which is likely to be an evolved star in each case.

\subsection{Other detected lines}
In addition to the main lines mentioned above, we made tentative detections in the \sio (1--0) $v=0$ and \choh 1$_1$--0$_0$ A$^{+}$ lines. The non-vibrationally excited state of \sio was detected towards G333.227$-$0.057, which is cospatial with a Class I \choh maser. This \sio transition is usually associated with outflows from regions of star formation. Given the above, a star formation origin of this \sio emission is likely. The \choh 1$_1$--0$_0$ A$^{+}$ and E lines are thermally emitting lines that are close in frequency (equivalent to 27.5\,\kms). We detect thermal emission at G333.227$-$0.057 and G333.314$+$0.109 in the A$^{+}$ transition, but not the E transition. \citet{kalenskii94} indicate that the E-\choh transition may require higher densities to produce line strengths comparable to A$^{+}$-\choh. The emission at G333.227$-$0.057 is cospatial with the thermal \sio source above and thermal methanol at G333.314$+$0.109 is cospatial with another Class I \choh maser, leading to a likely star formation origin for both thermal \choh sources.

\section{Discussion}
\subsection{Weather Effects}
As can be seen in Figure \ref{fig:cs}, significant ``striping'' artifacts exist in the data. The stripes coincide with the scanning direction of the observations. One contributing factor to the poor data is the fast-mosaic method; due to the fast transition times between pointings, some antennae were unable to keep up and did not contribute data. However, the main factor in poor data is the inclement weather experienced during observations. The most prominent stripes seen in Figure \ref{fig:cs} coincide with periods of light rain. Lagging antennae serve to decrease our sensitivity to emission, but the weather hampers our ability at determining real astrophysical emission from noise. This weather introduced noise into both broadband spectral windows, but has affected some frequencies more than others. See Figure \ref{fig:spectra} for \choh and \sio spectra. Examples of species with poor spectra include \choh 1$_1$--0$_0$ A$^{+}$ (and its E species), H53{\footnotesize{$\alpha$}} and H51{\footnotesize{$\alpha$}}. The noise is apparent on all relevant frequencies of emission, but fortunately has permitted reasonable detections for the primary lines of interest (\cs, Class I \choh and \sio masers).

It must be mentioned that the problems experienced in the pilot will not plague the overall MALT-45 survey, as the pilot survey merely seeks to provide a proof-of-concept on the techniques used and reveal potential problems before committing to the full observation.

\subsection{Telescope Performance}
\label{sec:telescope_performance}
Given the demands of our observational setup, we found some limitations to the ATCA that have affected the outcome of our pilot observations. Initially we found software limitations that meant changes to the intended survey design. These included: mosaic scans in Right Ascension and Declination coordinates only, rather than Galactic coordinates; no more than 999 pointings per mosaic file; no more than 2000 individual commands in a schedule file. These limitations resulted in a survey region that is at an angle to the Galactic plane, as can be seen in Figure \ref{fig:cs}, and a complicated mosaicing system.

During the observations, zoom modes with the CABB were not available, so we used the broadband mode. This had the unfortunate effect of greatly limiting the amount of spectral detail seen, as described in Section \ref{sec:pilotsurveyobs}. We also expect that the broadband mode is less sensitive to narrow-line emission (eg. masers) by a factor of up to three. Also as mentioned previously, we found that some of the antennae had difficulty in keeping up with the demanding 6-second pointing regime. Indeed we found that rms errors on the position of antenna 3 were typically 20 arcsec, which has reduced the sensitivity of the resulting data. We believe these excessive pointing errors can be eliminated with on-the-fly (OTF) mapping, where the antennae are driven smoothly and subsequent testing of this method suggests OTF mapping will resolve this issue for the full survey.

\subsection{Comparison with other observations}
\label{sec:comparison}
In this subsection, we compare selected sources within the MALT-45 pilot survey region to observations from previous surveys. In particular, we compare our results with the mid-infrared {\em Spitzer} Galactic Legacy Infrared Mid-Plane Survey Extraordinaire (GLIMPSE; \citealt{benjamin03}), for a visual correlation with infrared structures and objects. Mid-infrared extended emission is typically seen towards regions of active star formation, where the interstellar dust has been heated by newly formed stars. In addition to infrared emission, GLIMPSE also reveals prominent regions of infrared absorption, projected against the bright infrared background emission from the Galaxy. Such absorption features are referred to as IRDCs \citep{rathborne06}. These IRDCs are regions of high extinction where star formation may be about to take place in the near future, or has already taken place, but has not had sufficient time to break through the high extinction.

Infrared emission can also arise from a variety of other astrophysical objects. One such example is emission from evolved stars, that may also give rise to \sio masers, which appear as bright infrared stars in the GLIMPSE images \citep{robitaille08}. The evolved stars appear unresolved in the GLIMPSE images, which may occasionally be confused with confined sites of star formation. A useful, but not entirely robust, method to distinguish between evolved and young stars is that an infrared star is likely to be evolved if it is not associated with extended infrared emission or an IRDC and has a 4.5\,$\mu$m magnitude brighter than 7.8 \citep{robitaille08}. Young stars are more likely to be fainter and associated with either extended emission or IRDCs.

A comparison of MALT-45 \cs to HOPS \nh would be a valuable scientific addition to this paper, providing insight into star formation within these clouds. However, the data being presented is not of a sufficient quality to merit an accurate analysis. A variation in the \cs/\nh ratio may be apparent in a region such as (333.24, $+$0.21), where there appears to be an overabundance of \cs, relative to \nh, as can be seen in Figure \ref{fig:cs}. A thorough analysis will instead be performed with data from the full MALT-45 survey.\\

{\normalsize\em G333.227$-$0.057}\\
GLIMPSE images in Figures \ref{fig:glimpse} and \ref{fig:closeglimpse} reveal that the region at G333.227$-$0.057 is on the edge of extended emission, but does not appear to be closely associated with a prominent infrared feature. We note there is an infrared source located approximately 15 arcsec to the lower left of the maser position in Figure \ref{fig:closeglimpse}(d), which could be the powering source for an outflow that creates the maser. The Class I \choh maser spectrum (Figure \ref{fig:mopra_choh}) is dominated by a single peak at around -87\,\kms. This region has been studied previously, and contains OH maser emission \citep{sevenster97}, Class II \choh maser emission \citep{ellingsen96} and water maser emission \citep{kaufmann76,breen10}. The spectra of all maser species are remarkably similar, with a peak close to -87\,\kms, indicating a likely common origin for all masers. This velocity is significantly different from velocities of gas found associated with the G333 GMC, which are more typically around -45\,\kms. \citet{lockman79} calculate a distance of 3.6\,kpc to G333. Based on a kinematic model of Galactic rotation, we estimate the distance to G333.227$-$0.057 is 5.1\,kpc, putting it at a significant distance behind G333.

The Class I \choh maser associated with G333.227$-$0.057 was first detected by \citet{slysh94}, with a peak flux density over 300\,Jy. Using the high velocity resolution Mopra observations (Figure \ref{fig:mopra_choh}), we see that the spectrum shows weaker emission, with a peak flux density 250\,Jy, assuming a Jy/K conversion factor of 14.1 \citep{urquhart10}. This difference may be partially due to intrinsic variability of the maser, but can likely be attributed to calibration uncertainties which are at the 20\% level.

We have detected unresolved \sio (1--0) $v=0$ emission coincident with G333.227$-$0.057. This emission is seen in a single channel in the CABB data, but also as a well-defined single peak in the Mopra follow up data. Based on the Mopra spectrum at this position, the \sio emission has an integrated intensity of 2.2\,K\,\kms, a peak velocity of -88.6\,\kms and line full-width at half-maximum (FWHM) of 12.7\,\kms. The velocity of the \sio emission is within 2\,\kms of the peak of the Class I \choh maser, suggesting that the two are associated. As previously discussed, this ground state thermal transition is known to trace outflows associated with star formation. Using the CABB data, we also tentatively detect thermal \choh emission in the 1$_0$--0$_0$ A$^{+}$ transition. Since this transition was not covered by the Mopra follow up observations, we cannot be certain that this is a real detection. If this is a real detection, then we can expect occasional detection of these thermal lines throughout the main MALT-45 survey, which will be useful to compare the thermal and masing properties of \choh in such regions.\\

{\normalsize\em G333.726$+$0.366}\\
The \cs image shows a clear elongated structure at this position. Comparison with the GLIMPSE 8\,$\mu$m image indicates that there is an IRDC, whose morphology closely follows that of the \cs emission (Figure \ref{fig:closeglimpse}). HOPS \nh emission also closely follows the \cs emission and IRDC. The \nh emission is strongest at -34.9\,\kms and the \cs emission appears to peak close to this, although the velocity of the peak is not known accurately due to the broadband channels of the CABB. The close velocity and shape suggests both \cs and \nh arise from the IRDC. The IRDC is cospatial with bright, extended emission at 8\,$\mu$m which lies at the peak of both \cs and \nh emission. The extent of the \cs and \nh emission is almost 0.2$^{\circ}$, which, assuming a distance of 3.6\,kpc \citep{lockman79}, is equivalent to a projected distance of 10\,pc. The GLIMPSE image indicates that the IRDC may extend significantly beyond this.\\

{\normalsize\em G333.314$+$0.109}\\
We detect both \cs and a Class I \choh maser emission at this position. The GLIMPSE image of this region shows a faint IRDC, with an extended and reddened emission feature within the IRDC. The peak of weak \nh emission appears at this position, as well as a water maser, detected in HOPS. All spectral features appear to peak around -45\,\kms, which is similar to the velocity of the G333 GMC. It is likely that this cloud and the G333 GMC are located a similar distance from us.

\section{Future Work}
The pilot survey has been successful in many ways, perhaps most notably with the successful auto-correlation data reduction for each of the spectral lines relevant to MALT-45. The pilot has detected extended \cs emission, as well as eight Class I \choh masers, two \sio masers in the $v=1$ and $2$ transitions, and another tentative detection in the $v=2$ transition. We have also detected thermal emission in \cs, \sio $v=0$ and \choh 1$_0$--0$_0$ A$^{+}$ detections, with \cs emission being extended across a large fraction of the surveyed area. However, we anticipate great improvements for the main survey, based on the results of this work, and the expected completion of the CABB upgrade on the ATCA.

With CABB 64M-32k zoom modes available, we expect to achieve spectral channel resolution of $\sim$\,0.2\,\kms, which will not only allow us to identify more spectral features, but will also effectively increase our sensitivity to narrow-lined emission, such as from masers. For a maser of line 1\,\kms FWHM, we expect to gain a factor of $\sim$\,2.5 in sensitivity.

We expect to employ an OTF mapping technique, which will eliminate problems with the antenna drive systems not being able to keep up with the rigorous 6 second pointing we have used in this work. Early tests using OTF have indeed proven this to be the case. Overall, this could increase sensitivity by a factor of about 2, as some mapped regions had only three antennas sufficiently close to the desired pointing position to have usable data.

Early problems with the CABB blocks that cause noisy channels to be flagged out (as mentioned in Section \ref{sec:telescope_performance}) should be greatly reduced, as the cause of such problems become identified. We are currently aware that the problems with some blocks have already been addressed.

The ATCA scheduler now allows us to write schedule files in Galactic coordinates, as well as allowing many more pointings per mosaic and commands per schedule file. This will allow for a much smoother running schedule, as well as a smoother final mapped region, without jagged edges that would be a result of RA/Dec scans.

We will employ a method of scanning in two orthogonal directions (Galactic longitude and latitude). We expect that this will greatly reduce stripe features that are evident in the current data. Such stripes are mainly caused by varying weather conditions, but can also be affected by the changing elevation of the observations. Our experience in the pilot work is that observations should not be undertaken under heavy cloud, light rain or worse weather conditions. Scanning in two directions has proven very successful for OTF mapping with the Mopra radiotelescope, such as those used in HOPS. Scanning in two directions will also increase the sensitivity. Furthermore, the combination of both OTF mapping and orthogonal scans will produce data that is effectively Nyquist sampled with closest spacing between adjacent observations of 30 arcsec. Overall, we expect the sensitivity to masers in the full MALT-45 survey to be around 0.1\,Jy. The differences between both surveys are highlighted in Table \ref{tab3}.

\begin{table*}
  \begin{center}
  \caption{MALT-45 parameters from the pilot survey and full survey.}
  \label{tab3}
  \begin{tabular}{ l c c c c c c}
    \hline
    MALT-45 & Correlator & Channel resolution & Cycle time & Targeted & On-the-fly & Orthogonal scans/ \\
            & configuration & (\kms)          & (seconds)  & spectral lines & mapping & Nyquist sampling \\
    \hline
    Pilot survey & 2$\times$ 2048\,MHz & $\sim$7 & 6 & 12 & No & No \\
                 & broadband windows & & & & & \\
    Full survey & 32$\times$ 64\,MHz   & $\sim$0.2 & 6 & 12 & Yes & Yes \\
                & zoom windows & & & & & \\
    \hline
  \end{tabular}
  \end{center}
\end{table*}

\section*{Acknowledgments}
The authors acknowledge and thank the referee, J. Martin-Pintado, for their comments that have improved the quality of this paper. We thank CSIRO Astronomy and Space Science for allowing the use of the ATCA in this project, and for the development of CABB. We thank the staff at Narrabri for their assistance and testing of the small mosaic transition timings on the ATCA before the pilot survey observations. NL acknowledges partial support from the ALMA-CONICYT Fund for the Development of Chilean Astronomy Project 31090013, Center of Excellence in Astrophysics and Associated Technologies (PFB 06) and Centro de Astrof\'{i}sica FONDAP\,15010003. The Australia Telescope Compact Array and the Mopra radio telescope are part of the Australia Telescope National Facility which is funded by the Commonwealth of Australia for operation as a National Facility managed by CSIRO.

\label{lastpage}
\end{document}